\documentclass{elsart}
\usepackage{graphicx}
\usepackage[fleqn]{amsmath}
\usepackage{amssymb}
\newcommand{\be}{\begin{equation}}
\newcommand{\ee}{\end{equation}}
\newcommand{\bea}{\begin{eqnarray}}
\newcommand{\eea}{\end{eqnarray}}

\newcommand{\sla}[1]{\ooalign{\hfil/\hfil\crcr$#1$}}

\allowdisplaybreaks[2]

\begin{document}
\begin{frontmatter}

\title{On the finite size mass shift formula \\
for stable particles}

\author{Yoshiaki Koma and Miho Koma}~\footnote{present 
address:
DESY, Theory Group, 
Notkestrasse 85, D-22603 Hamburg, Germany}

\address{Max-Planck-Institut f\"ur Physik,
F\"ohringer Ring 6, D-80805, M\"unchen, Germany}

\begin{abstract}
L\"uscher's finite size mass shift formula 
in a periodic finite volume, involving
forward scattering amplitudes
in the infinite volume, is revisited
for the two stable distinguishable particle system.
The generalized mass shift formulae 
for the boson and fermion are
derived in the boson-boson and fermion-boson
systems, respectively.
The nucleon mass shift formula is given in the 
nucleon-pion system and the relation to the computation
within chiral perturbation theory
is discussed.
\end{abstract}

\end{frontmatter}

\section{Introduction}
\label{sec:intro}

\par
Nowadays the control of finite volume effects 
in lattice QCD simulations with dynamical fermions
becomes a more important issue in order to 
determine the hadron spectrum precisely.
Applications of chiral perturbation theory
(ChPT)~\cite{Becher:1999he} to the measured
spectrum on the lattice have been done 
not only to achieve the chiral 
extrapolation~\cite{Bernard:2003rp,Procura:2003ig},
but also to find its finite volume dependence
towards the thermodynamic 
limit~\cite{Orth:2003nb,Colangelo:2003hf,%
AliKhan:2003cu,Beane:2004tw,Beane:2004ks}.

\par
In this context, L\"uscher's formula, relating the 
mass shift in finite volume with periodic boundary 
conditions to forward elastic scattering amplitudes 
in infinite volume, provides us with an elegant tool for
such a purpose~\cite{Luscher:1983rk}.
In the latter reference L\"uscher presented
a rigorous proof of his formula for the case of a 
self-interacting boson to all orders in perturbation 
theory~\cite{Luscher:1986dn}.
By writing the difference of the masses between
in finite and infinite volumes as $\Delta m(L) =M(L)-m$
with the spatial box of the size $L^{3}$,
the formula is given by
\bea
\Delta m (L) 
= 
- \frac{3}{8 \pi m L}
\Biggl [ 
\! 
\frac{\lambda^{2}}{4 \nu_{B}}e^{-\frac{\sqrt{3}}{2}Lm}
\! + \! \! 
\int_{-\infty}^{\infty} \! 
\frac{dq_{0}}{2\pi} 
e^{-L\sqrt{q_{0}^{2}+m^{2}}} F(iq_{0})
\!
\Biggr ] \! \! + \! O(e^{-\sqrt{3/2}L m}) 
\; , \quad
\label{eqn:Luescher-boson-massshift}
\eea
where $F(\nu)$ is the 
forward scattering amplitude ($\nu=iq_{0}$ denotes the
crossing variable), which has poles at
$\nu= \pm \nu_{B}$ with $\nu_{B}= m/2$. 
The coupling $\lambda$ is then defined by the relation
\bea
\lim_{\nu \to \pm \nu_{B}}(\nu^{2}-\nu_{B}^{2})
F(\nu) =\frac{\lambda^{2}}{2}\; .
\eea

\par
One may ask if one can immediately generalize this formula
to a theory describing the interaction of two stable
particles $A$ and $B$ (or multiplets thereof)
of different masses, $m_{A}$ and $m_{B}$, respectively.
One may find that the r.h.s. of 
Eq.~\eqref{eqn:Luescher-boson-massshift} can be 
expressed by using only the second term
if the integral path is extended to
the complex $q_{0}$ plane such as
$-\infty+i \zeta  \to +\infty+ i\zeta$
with $m/2 < \zeta < m$.
In fact, the shift of the integral path back to 
the real $q_{0}$ axis ($-\infty  \to +\infty$)
picks up a residue associated with
$F(\nu)$ at the pole
$\nu = + \nu_{B}$, which leads to the first term of 
Eq.~\eqref{eqn:Luescher-boson-massshift}.
In this sense, we may call the first term 
in Eq.~\eqref{eqn:Luescher-boson-massshift} 
{\em the pole term}.
Then, we may speculate that the asymptotic 
formula for the $A$-particle mass shift for the
case $m_{A}\gg m_{B}$ is written as
\bea
\Delta m_{A}  
&\stackrel{?}{\approx}&
- \frac{3}{8 \pi m_{A} L}
\int_{-\infty+i\zeta}^{\infty+i\zeta} \frac{dq_{0}}{2\pi} \; 
e^{-L\sqrt{q_{0}^{2}+m_{B}^{2}}} F_{\mathit{AB}}(iq_{0}) 
\nonumber\\*
&=&
- \frac{3}{8 \pi m_{A} L} \Biggl [
\frac{\lambda^{2}}{4 \nu_{B}}e^{-L \sqrt{m_{B}^{2}-\nu_{B}^{2}}}
+\int_{-\infty}^{\infty} \frac{dq_{0}}{2\pi} \; 
e^{-L\sqrt{q_{0}^{2}+m_{B}^{2}}} F_{\mathit{AB}}(iq_{0}) 
\Biggr ] \; ,
\label{eqn:formula-doubt}
\eea
with $F_{\mathit{AB}}(\nu)$,
the forward scattering
amplitude for the $A+B \to A+B$ process,
and in this case $\nu_{B}=m_{B}^{2}/2m_{A}$.
For instance, by identifying $A$ with the nucleon 
and $B$ with the pion,
and exploiting the residue of $F_{\mathit{N\pi}}(\nu)$ 
at $\nu= +\nu_{B}$, 
one would then reproduce the nucleon mass shift formula
presented by L\"uscher in his Cargese 
lectures~\cite{Luscher:1983rk}.

\par 
Among the applications of ChPT, the QCDSF-UKQCD collaboration
estimated the finite volume effect on the nucleon mass from data in
$N_{f}=2$ lattice QCD~\cite{AliKhan:2003cu}, applying the mass shift
formula derived within ChPT. 
Along the way, they found however, when
expressing their formula in terms of $F_{N \pi}(\nu)$ in the same
order of ChPT, that the factor of the pole term is twice larger than
that of L\"uscher's in Ref.~\cite{Luscher:1983rk}, although the rest
is the same.  There seems to be no mistake in the formula in
Ref.~\cite{AliKhan:2003cu}, at least within the infrared
regularization scheme~\cite{Becher:1999he}.  
On the other hand,
L\"uscher's formula being considered as general such that it can also
be applicable to ChPT at any orders, this discrepancy poses a
structural question.

\par 
In this paper, we thus investigate the mass shift formula for the
interacting two stable particle system along the lines of L\"uscher's
proof for a self-interacting bosonic theory.  We will not assume
$m_{A} \gg m_{B}$ (as in the physical case above) because this would
obscure the general structure of the formula, so that we will keep
terms which would be relevant if the masses were close.  In
Sect.~\ref{sec:result}, we first 
describe the result on the generalized mass
shift formulae for the boson and fermion in the boson-boson and
fermion-boson systems, respectively.  In Sect.~\ref{sec:derivation},
we then provide a part of the derivation by selecting a typical diagram for
the self energy in the two particle system.  
We discuss the nucleon mass shift formula in the $N$-$\pi$ system in
Sect.~\ref{sec:nucleon}.  
A summary is given in Sect.~\ref{sec:summary}.  
Notations used in this paper are summarized
in Appendix~\ref{sec:notation}.

\section{Asymptotic formulae for finite size mass shift}
\label{sec:result}

\par
In this section, we describe the result
on the generalized mass shift formulae for the boson 
and fermion in the boson-boson and fermion-boson
systems, respectively.

\par
The physical mass of a stable particle is given by the position
of the pole of the propagator of an asymptotic field.
In the framework of perturbation theory the pole is
shifted from the bare one due to the self energy
arising from polarization effects from the virtual particles.
In finite volume, the expressions for the self energy involve sums
over discrete spatial loop momenta, 
$\vec{q}(L) = 2\pi \vec{n}/L$ $(\vec{n} \in \mathbb{Z}^{3})$,
instead of integrals.
By using the Poisson summation formula, 
such a summation can be rewritten again as
an integral with another summation over integer vectors 
$\vec{m} \in \mathbb{Z}^{3}$
and an exponential factor depending on $L$:
\bea
\frac{1}{L^{3}} \sum_{\vec{n} \in \mathbb{Z}^{3}}
\int_{}^{}\frac{dq_{0}}{2 \pi}  f (q_{0},\vec{q}(L) )
=
\sum_{\vec{m} \in \mathbb{Z}^{3}}^{} 
\int_{}^{}\frac{d^{4}q}{(2 \pi)^{4}} e^{-iL \vec{m}\cdot \vec{q}}
f (q_{0},\vec{q} \; )\; ,
\label{eqn:poisson}
\eea
where $f (q)$ is a function composed of
propagators and vertex functions.
Then, the difference of the self energies
between the finite and infinite volumes,
appearing in the definition of the mass shift 
as described below, 
can be defined by the sum over $|\vec{m}| \ne 0$, since 
$|\vec{m}|=0$ corresponds to the integral in infinite
volume~\cite{Hasenfratz:1990pk}.
The asymptotic formula at large $L$, which is of our interest, is
given by the contribution of~$|\vec{m}| =1$.
(next leading contribution is $|\vec{m}| =\sqrt{2}$.)
In this case, the exponential factor in Eq.~\eqref{eqn:poisson}
is reduced to $2 \sum_{i=1}^{3} \cos ( L q_{i})$.

\subsection{Boson mass shift formula in the 
$\phi_{A}$-$\phi_{B}$ system}

\begin{figure}[t]
\centering\includegraphics[width=13.5cm]{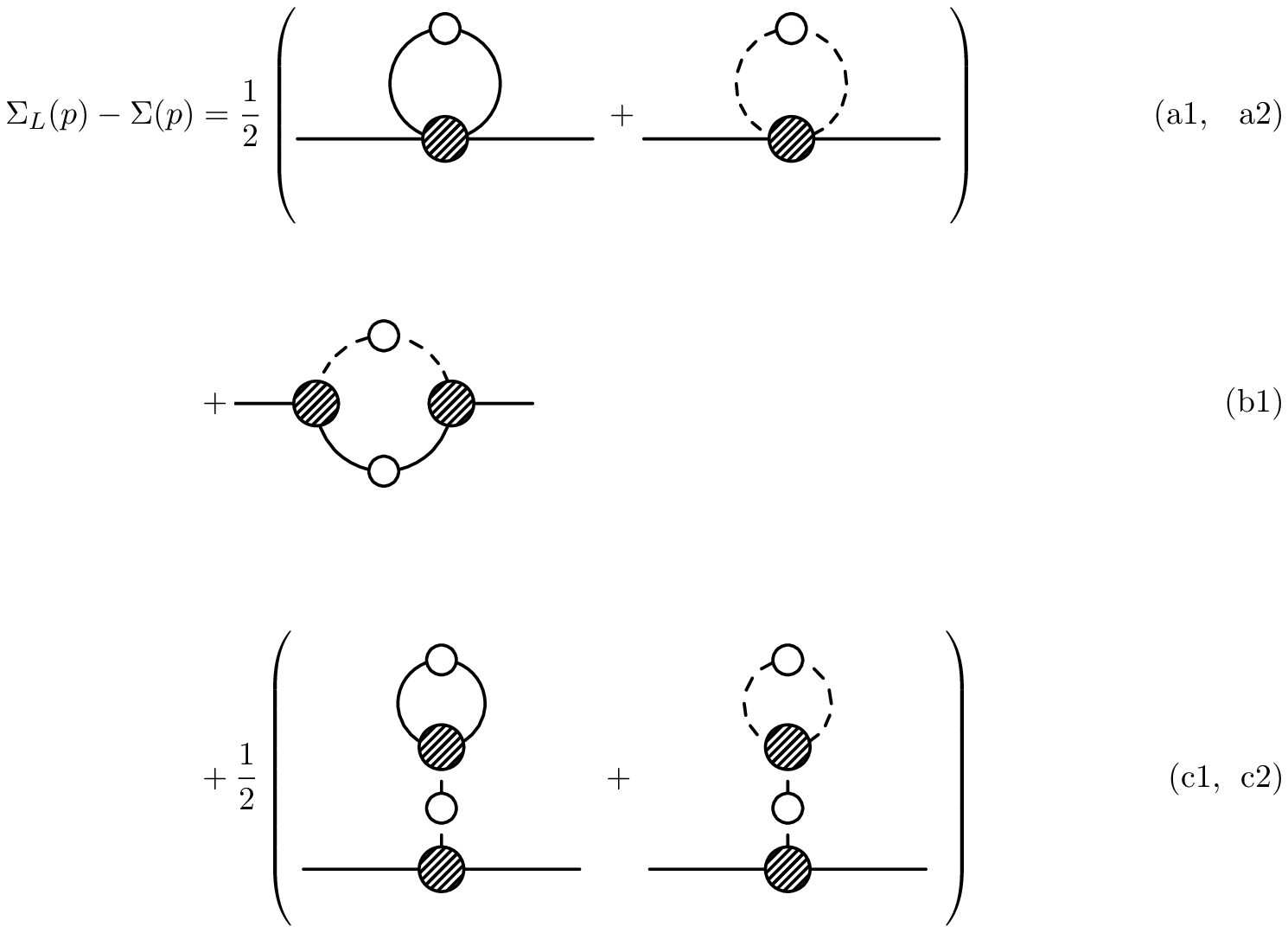}
\caption{Self-energy diagrams which contribute to the mass 
shift formula  in the  $\phi_{A}$-$\phi_{B}$  system. 
A solid line with an empty circle 
corresponds to the propagator of $\phi_{A}$ 
and a dashed line to that of $\phi_{B}$. 
A shaded blob is a vertex function, see Appendix~\ref{sec:notation}
for the definition.
It is assumed that $\phi_{A}$ carries a conserved charge.} 
\label{fig:selfenergy}
\end{figure}

\begin{figure}[t]
\centering\includegraphics[width=13.5cm]{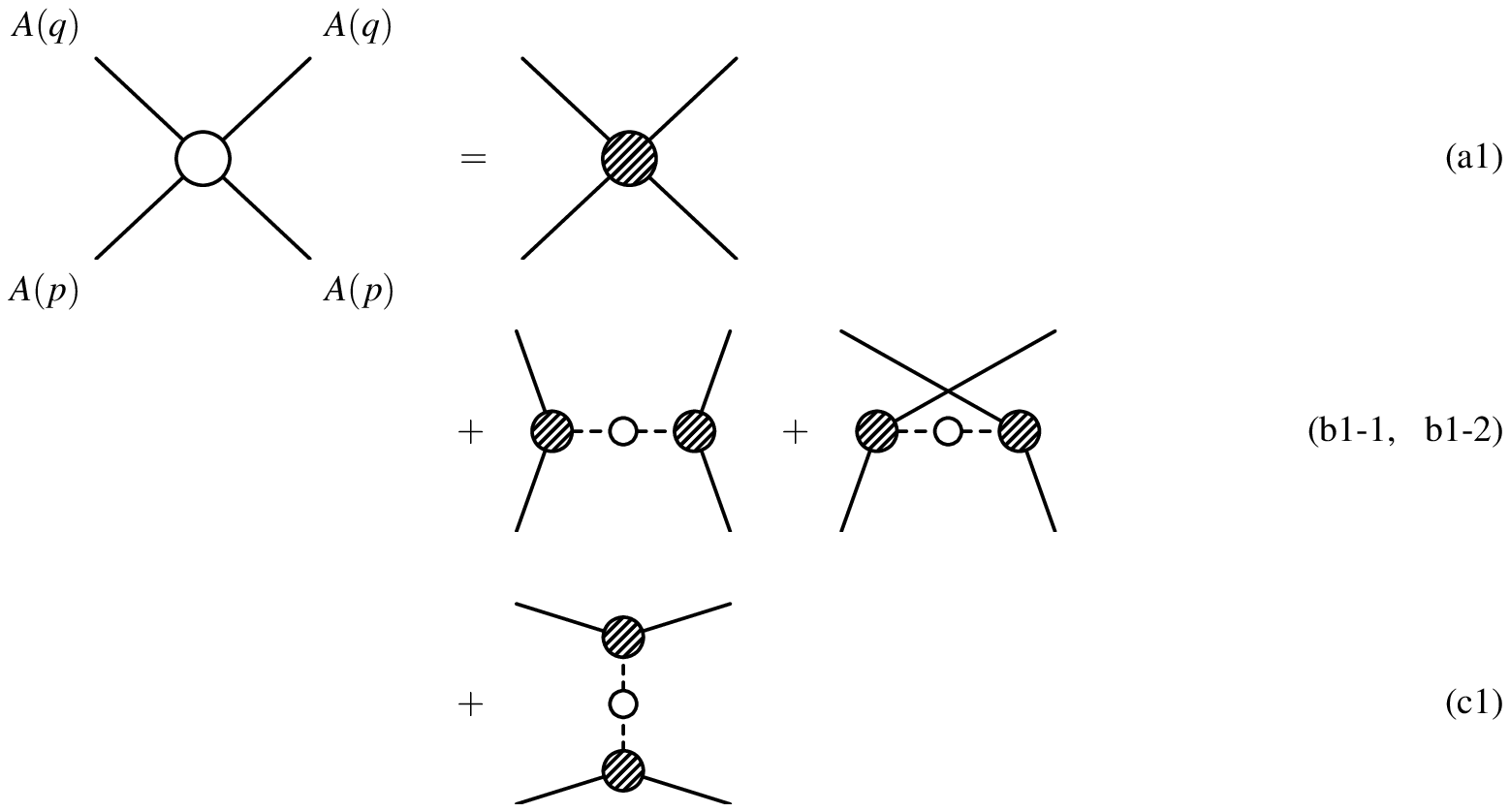}
\caption{Ingredients of $F_{\mathit{AA}}(\nu)$ in the 
$\phi_{A}$-$\phi_{B}$ system.
The labels represent the correspondence with
the self-energy diagrams in Fig.~\ref{fig:selfenergy}.
The diagram (b1) in Fig.~\ref{fig:selfenergy} is responsible for
the two types of amplitudes.}
\label{fig:aa-scattering}
\end{figure}

\par
\begin{figure}[t]
\centering\includegraphics[width=13.5cm]{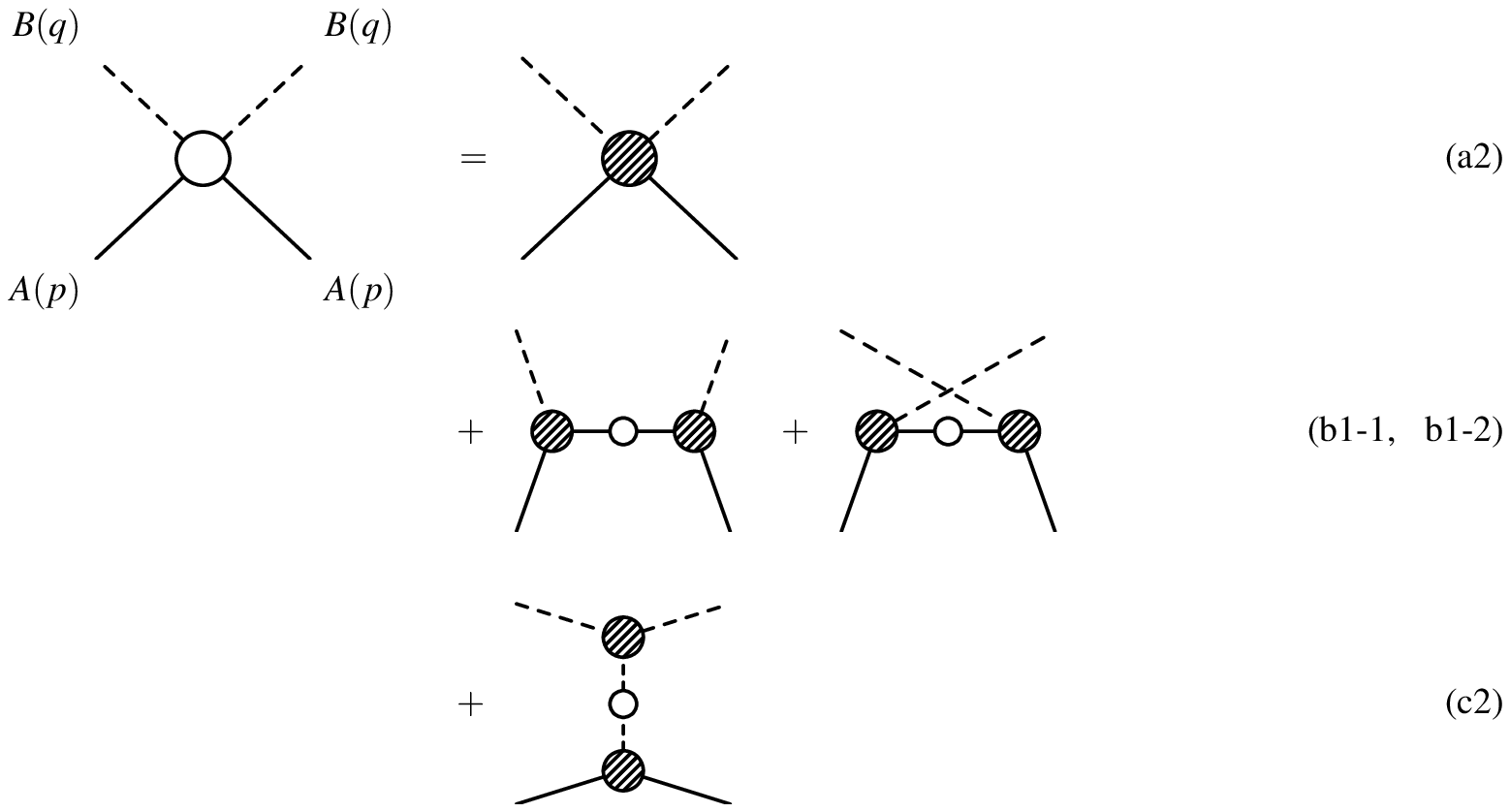}
\caption{Ingredients of $F_{\mathit{AB}}(\nu)$
in the $\phi_{A}$-$\phi_{B}$ system
(or in the $\Psi_{A}$-$\phi_{B}$ system 
with arrows to the solid lines).
The labels represent the correspondence with 
the self-energy diagrams in Fig.~\ref{fig:selfenergy}.}
\label{fig:ab-scattering}
\end{figure}

As the finite and infinite volume
masses $M_{A}(L)$, $m_{A}$ of $A$-particle are 
defined from the poles of the propagators of $\phi_{A}$ 
(see Appendix~\ref{sec:notation}),
$\Delta m_{A} (L) = M_{A} (L) -m_{A}$
is reduced to
\bea
\Delta m_{A} (L) = 
- \frac{1}{2m_{A}}[\Sigma_{L}(p)-\Sigma(p)]
+O ( (\Delta m_{A})^{2})
\quad \mbox{at $p=(im_{A},\vec{0} \; )$} \; ,
\label{eqn:massshift-2boson}
\eea
where $\Sigma_{L}(p)$ and $\Sigma(p)$ stand for the self 
energies of $\phi_{A}$ in the finite and infinite volumes, 
respectively.
We work in Euclidean space as in Ref.~\cite{Luscher:1986dn}.
Here, since the fermion mass shift formula in the fermion-boson 
system is of our  final interest,
we assume that only $\phi_{A}$ carries a conserved charge
so that interaction induced only by the 3-point vertex 
$\mathit{AAB}$ and 4-point charge conserving vertices are 
taken into account.
In such a system, the difference of the self energy of $\phi_{A}$ 
in the finite and infinite
volumes at any orders in perturbation theory 
is provided by the diagrams listed in Fig.~\ref{fig:selfenergy}.
For $m_{B}/m_A \in (0,1]$,
Eq.~\eqref{eqn:massshift-2boson} is reduced to
\bea
\Delta m_{A} (L)
&=&
 -  \frac{3}{8\pi m_{A} L }
\Biggl [
\frac{ \lambda_{\mathit{AAB}}^{2}}{ 2 \nu_{B}}
e^{- L \sqrt{m_{B}^{2}-\nu_{B}^{2}}}
+ \int_{-\infty}^{\infty} \frac{dq_{0}}{2 \pi}
\; e^{-L\sqrt{q_{0}^{2}+m_{A}^{2}}} F_{\mathit{AA}}(i q_{0})
\nonumber\\*
&&
+ \int_{-\infty}^{\infty}\frac{dq_{0}}{2 \pi}
\; e^{-L \sqrt{q_{0}^{2}+m_{B}^{2}}} F_{\mathit{AB}}(iq_{0})
\Biggr ] 
+ O( e^{- L\bar{m}}) \; ,
\label{eqn:massshift-boson}
\eea
where $F_{\mathit{AB}}(\nu)$ and $F_{\mathit{AA}}(\nu)$ 
denote the forward scattering amplitudes of the processes
$\phi_{A}(p)+\phi_{B}(q) \to \phi_{A}(p)+\phi_{B}(q)$ 
and $\phi_{A}(p)+\phi_{A}(q) \to \phi_{A}(p)+\phi_{A}(q)$ 
in the infinite volume, respectively.
The ingredients of these scattering amplitudes
are graphically represented in 
Figs.~\ref{fig:aa-scattering} and~\ref{fig:ab-scattering}.
$\lambda_{\mathit{AAB}}$ 
is an effective renormalized coupling defined from the residue 
of $F_{\mathit{AB}}(\nu)$ at $\nu= \pm \nu_{B} 
= \pm m_{B}^{2}/2m_{A}$ as
\bea
\lim_{\nu \to \pm \nu_{B}}(\nu^{2}-\nu_{B}^{2}) \; 
F_{\mathit{AB}}(\nu)
=\frac{\lambda_{\mathit{AAB}}^{2}}{2}\; .
\label{eqn:eff-re-coupling}
\eea
The pole term in Eq.~\eqref{eqn:massshift-boson}
is associated with the evaluation of the self-energy 
diagram~(b1) in Fig.~\ref{fig:selfenergy}.
The error term is defined by
\bea
\bar{m} \ge 
\sqrt{2(m_{B}^{2}-\nu_{B}^{2})}\; ,
\label{eqn:error}
\eea 
which is due to the neglect of~$|\vec{m}| \geq \sqrt{2}$ 
contributions.
If $m_{B}/m_{A} < \alpha_{c} =\sqrt{2-\sqrt{2}} \approx 0.765$,
we may neglect the second term in 
Eq.~\eqref{eqn:massshift-boson},
which decays more rapidly than the error term at large $L$.
Note that the pole term in Eq.~\eqref{eqn:massshift-boson}
is twice larger than that in Eq.~\eqref{eqn:formula-doubt}
and two types of forward scattering amplitudes
contribute to the formula 
in Eq.~\eqref{eqn:massshift-boson}.

\subsection{Fermion mass shift formula in the 
$\Psi_{A}$-$\phi_{B}$ system}

\begin{figure}[t]
\centering\includegraphics[width=13.5cm]{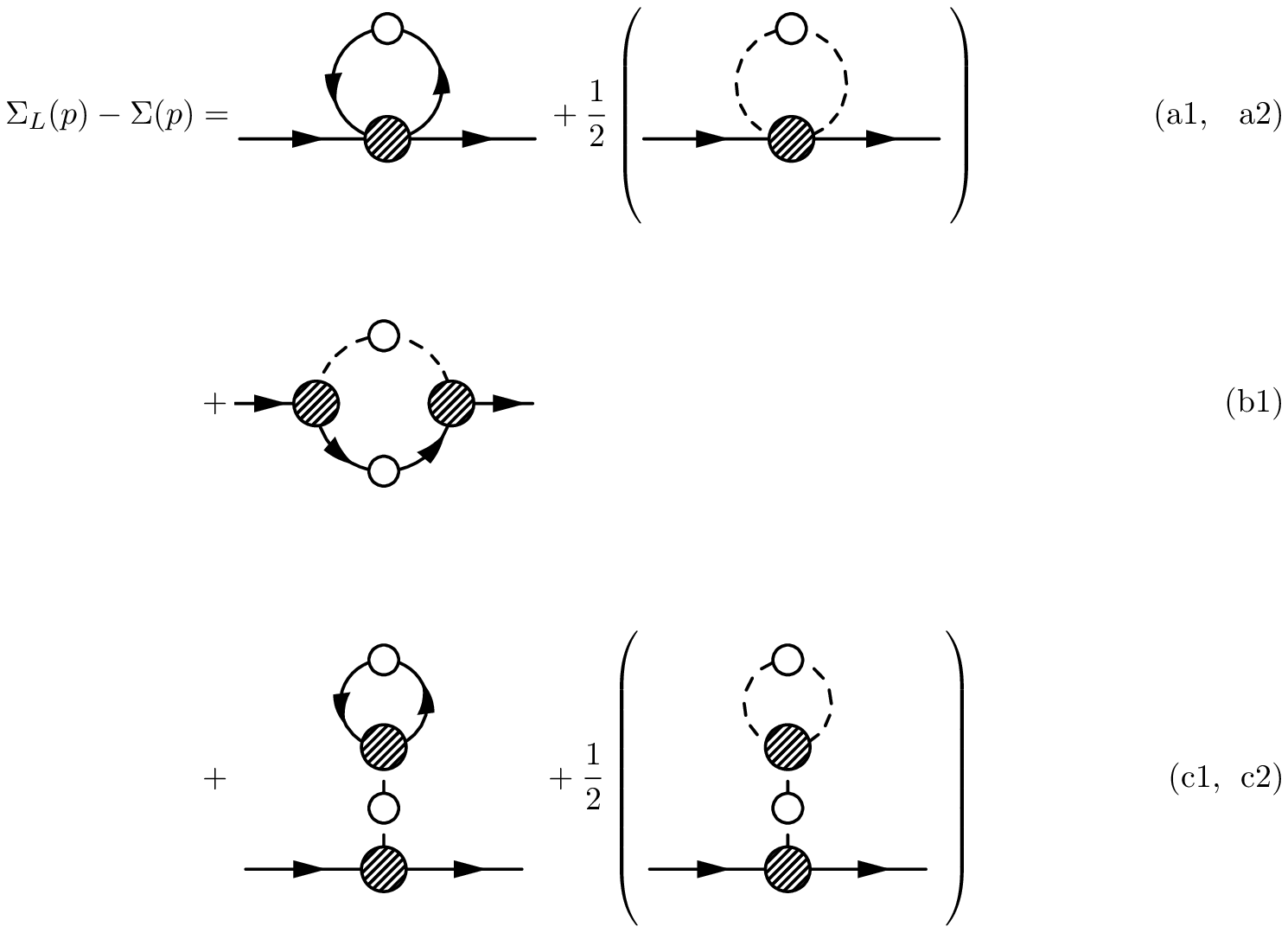}
\caption{Self-energy diagrams which contribute to the mass shift
formula in the $\Psi_{A}$-$\phi_{B}$ system. 
A solid arrow with an empty circle
corresponds to the propagator of $\Psi_{A}$.}  
\label{fig:selfenergy-fermion}
\end{figure}

\begin{figure}[t]
\centering\includegraphics[width=13.5cm]{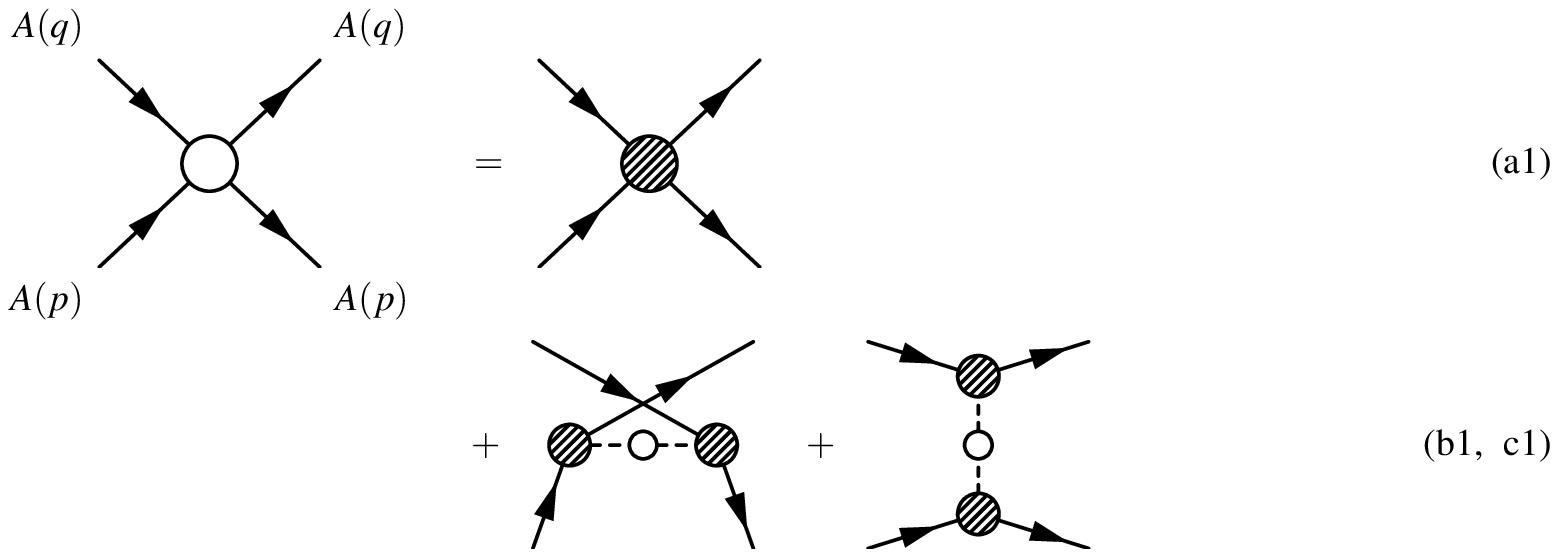}
\caption{Ingredients of $F_{\mathit{AA}}(\nu)$ in 
the  $\Psi_{A}$-$\phi_{B}$ system. 
The labels represent the correspondence with
the self-energy diagrams in Fig.~\ref{fig:selfenergy-fermion}.}
\label{fig:aa-fermion-scattering}
\end{figure}

\begin{figure}[t]
\centering\includegraphics[width=13.5cm]{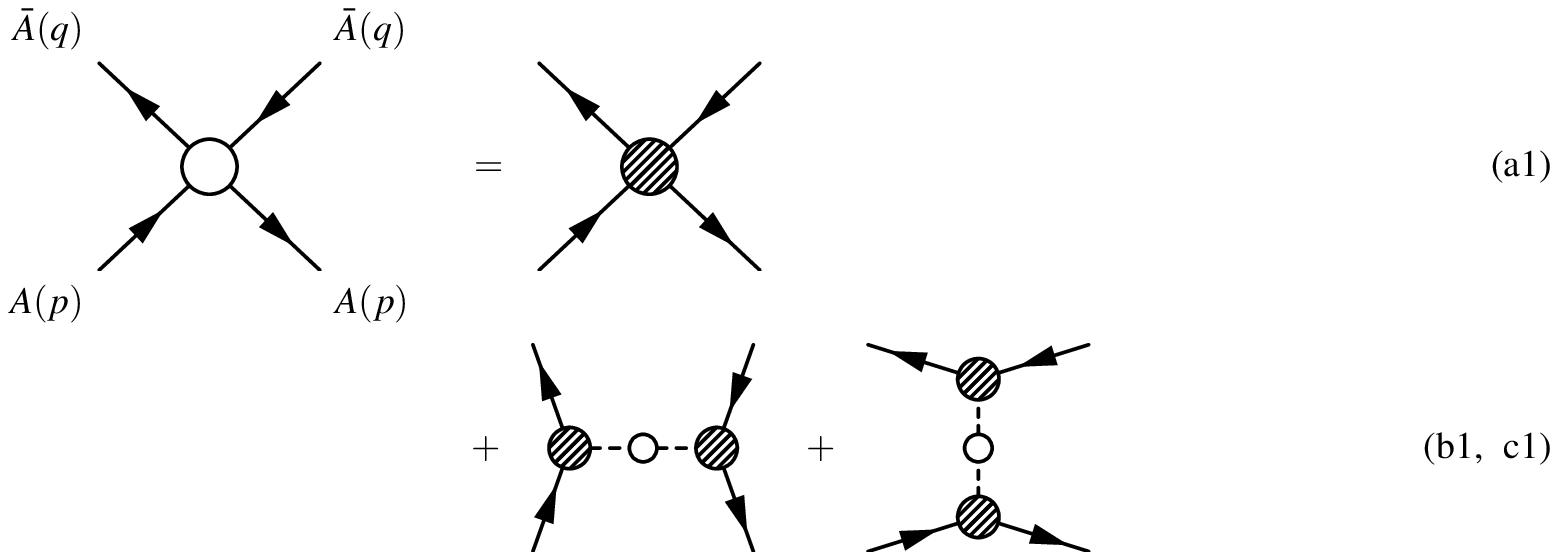}
\caption{Ingredients of $F_{\mathit{A\bar{A}}}(\nu)$ in 
 the $\Psi_{A}$-$\phi_{B}$ system. 
The labels represent the correspondence with
the self-energy diagrams in Fig.~\ref{fig:selfenergy-fermion}.}
\label{fig:aabar-fermion-scattering}
\end{figure}

From the poles of the propagators of $\Psi_{A}$ in the
finite and infinite volumes (see Appendix~\ref{sec:notation}), 
we obtain
\bea
\Delta m_{A} = 
-
\frac{1}{2m_{A}}\bar{u}
(p,r)[\Sigma_{L}(p)-\Sigma(p)]u(p,r)
+ O ((\Delta m_{A})^{2} )
\;\; \mbox{at $p=(i m_{A},\vec{0} \; )$}  \; , \;\; \;
\label{eqn:massshift-fermion-boson}
\eea
where $\bar{u}(p,r)$ and 
$u(p,r)$ are spinors of $\Psi_{A}$ with the spin $r$.
The mass shift is then written down as 
\bea
\Delta m_{A} (L)
&=&
-
\frac{3}{8\pi m_{A} L}
\Biggl [
\frac{\lambda_{\mathit{AAB}}^{2}}{2 \nu_{B}}
e^{- L\sqrt{m_{B}^{2}-\nu_{B}^{2}}}
\nonumber\\*
&&
- \int_{-\infty}^{\infty} \frac{dq_{0}}{2 \pi}
\; e^{-L\sqrt{q_{0}^{2}+m_{A}^{2}}}
\{ F_{\mathit{AA}}(iq_{0})+F_{\mathit{A\bar{A}}}(iq_{0}) \}
\nonumber\\*
&&
+
\int_{-\infty}^{\infty} \frac{dq_{0}}{2 \pi}
\; e^{-L\sqrt{q_{0}^{2}+m_{B}^{2}}} F_{\mathit{AB}}(iq_{0})
\Biggr ] 
+ O(e^{-L\bar{m}}) \; ,
\label{eqn:massshift-fermion}
\eea
where $\bar{m}$ is the same as in Eq.~\eqref{eqn:error}.
Now, three types of the forward scattering amplitudes 
contribute to the formula,
$F_{\mathit{AB}}(\nu)$: 
$\Psi_{A}(p)+\phi_{B}(q) \to  \Psi_{A}(p)+\phi_{B}(q)$ (Compton type),
$F_{\mathit{AA}}(\nu)$: 
$\Psi_{A}(p)+\Psi_{A}(q) \to  \Psi_{A}(p)+\Psi_{A}(q)$ 
(M$\sla{\rm o}$ller type),
and
$F_{\mathit{A\bar{A}}}(\nu)$: 
$\Psi_{A}(p)+\bar{\Psi}_{A}(q) \to \Psi_{A}(p)+\bar{\Psi}_{A}(q)$
(Bhabha type) (see, 
Figs.~\ref{fig:ab-scattering},~\ref{fig:aa-fermion-scattering} 
and~\ref{fig:aabar-fermion-scattering}).
Again we may neglect the second term if $m_{B}/m_{A}<\alpha_{c}$.
Apart from the relative minus sign in front of the second
term, which is due to Fermi statistics, the 
formula is almost the same as for the boson.
The factor of the pole term is twice larger
than that in Eq.~\eqref{eqn:formula-doubt}.


\section{Outline of derivation}
\label{sec:derivation}

\par
In this section, we provide a part of the 
derivation of the mass shift formulae in
Eqs.~\eqref{eqn:massshift-boson} and~\eqref{eqn:massshift-fermion}
for the case 
\bea
\alpha \equiv \frac{m_{B}}{m_{A}} \quad  \in (0,1] \; .
\eea
We shall here concentrate on the evaluation of the 
self-energy diagram~(b1) in Figs.~\ref{fig:selfenergy} 
and~\ref{fig:selfenergy-fermion},
which is typical in the two particle system and
finally provides the pole term $O(e^{-L\mu})$ with
\bea
\mu \equiv  m_{B} \sqrt{1-\frac{\alpha^{2}}{4}}
=\sqrt{m_{B}^{2}-\nu_{B}^{2}}\;,
\qquad
\nu_{B}=\frac{m_{B}^{2}}{2m_{A}}
\; .
\eea
The definitions of the propagators and the vertex functions
used in the following calculations
are summarized in Appendix~\ref{sec:notation}.
Other self-energy diagrams can be evaluated in a similar way.
The results are collected in Appendix~\ref{sec:summary-result}.

\subsection{Boson mass shift formula}

The self-energy diagram (b1) in 
Fig.~\ref{fig:selfenergy} with $|\vec{m}|=1$
(see Eq.~\eqref{eqn:poisson}) is given by
\bea
I_{b1} &=& 
\int  \!\! \frac{d^{4}q}{(2 \pi)^{4}} \!\!
\left ( 2 \sum_{j=1}^{3} \cos L q_{j}\right )
\Gamma_{\mathit{AAB}}(-p, (1-\delta)p+q ;  \delta p-q) 
G_{A}( (1-\delta)p+q) 
 \nonumber\\*
&& \times 
G_{B}( \delta p - q) 
\Gamma_{\mathit{AAB}}
(p, -(1-\delta)p-q ; -\delta p + q)  \biggr |_{p=(im_{A},
\; \vec{0} \, )}\; ,
\label{eqn:integral-b1}
\eea 
where $\delta \in (0,\frac{1}{2}]$
is a parameter chosen optimally below.
We are working in Euclidean space so that
$q_{\mu}=q^{\mu}=(q_{0},\vec{q}~)$ and
$\vec{q}=(q_{1},q_{2},q_{3})$.
Inner product is defined by $pq=p_{0}q_{0}+\vec{p}\cdot 
\vec{q}=im_{A}q_{0}$.
Due to rotational invariance among 
$q_{1}$, $q_{2}$ and $q_{3}$ one can 
rewrite the cosine factor in Eq.~\eqref{eqn:integral-b1} as
$2 \sum_{j=1}^{3} \cos L q_{j} \to 6 \; \mathrm{Re}\; e^{i L q_{1}}$.

\par
Firstly, we integrate out $q_{1}$.
To do this, we extend $q_{1}$ to a complex variable
and perform complex contour integration.
Here, the integral path must be chosen appropriately
by considering the analyticity properties of 
the integrand, the vertex functions and the propagators,
in the complex $q_{1}$ plane.

\par
A complex domain $\mathbb{D}=\{(p,q) \in \mathbb{C}^{4} 
\times \mathbb{C}^{4}\}$,
to which the vertex function
$\Gamma_{\mathit{AAB}}
(\mp p, \pm (1-\delta)p \pm q; \pm \delta p  \mp q )$
initially defined for $(p,q) \in \mathbb{R}^{4}
\times \mathbb{R}^{4}$
is  analytically extended, can be found as follows.
The basic assumption here is that 
the vertex function at any orders in perturbation theory
consists of a set of $A$ and $B$ lines (free propagators).
The $l$-th $A$ or $B$ line
is then parametrized as
$[(k(l)+r(l))^{2}+m_{A}^{2}]^{-1}$ or 
$[(k(l)+r(l))^{2}+m_{B}^{2}]^{-1}$, 
where $k(l)$ is the flow of
external momentum given by a combination of
complex variables $p$ and $q$, and $r(l)$ is 
a combination of internal loop momenta to be integrated out, 
which is a real variable in Euclidean space.
Thus, we find that the vertex function has no singularity if 
$(\mathrm{Im}~k(l))^{2} < m_{A}^{2}$ 
and $(\mathrm{Im}~k(l))^{2} < m_{B}^{2}$
are satisfied for all $A$ and $B$ lines.
In order to find the possible choices of $k(l)$,
we label the three bare vertices where
the external momenta,
$\mp p$, $\pm (1-\delta)p \pm q$ and $\pm \delta p  \mp q$,
are plugged in
as $a_{1}$, $a_{2}$ and $b$, respectively
(e.g. $a_{1}=a_{2}=b$ at the tree level).
We consider here the case that $A$-particle carries a
conserved charge.
In this case, there always exists a set of $A$ lines 
connecting $a_{1}$ and $a_{2}$.
We can then choose $k(l)$
up to overall sign as follows.
If the connected $A$ lines flow through $b$, 
such $A$ lines carry $(1-\delta) p - \frac{1}{2}q$ 
between $a_{1}$ and $b$, and
$(1-2 \delta) p + \frac{1}{2}q$
between $b$ and $a_{2}$.
The other lines then carry $\delta p + \frac{1}{2}q$ or $0$.
If the connected $A$ lines do not flow through $b$, 
$A$ lines carry $(1-\delta) p + \frac{1}{2}q$ 
between $a_{1}$ and $a_{2}$, and
the other lines carry $\delta p$, $\frac{1}{2}q$,
$\delta p - \frac{1}{2}q$ or $0$.
The analytic complex domain $\mathbb{D}$ is
then specified by inserting these $k(l)$ 
into the above two inequalities.
Practically, it is enough to consider the cases
\bea
&&
\left (
\mathrm{Im} \{ (1-\delta)\, p \pm \frac{1}{2} \,q \} 
\right )^{2} < m_{A}^{2}\;,
\label{eqn:ineq1}
\\
&&
\left(\mathrm{Im} \{ \delta\, p \pm \frac{1}{2} \,q \} 
\right)^{2} < m_{B}^{2}\; ,
\label{eqn:ineq2}
\eea
so that all other cases are fulfilled.~\footnote{In a 
similar way, one can discuss
the analyticity properties of $\Gamma_{\mathit{AAB}}$
for the case neither $\phi_{A}$ nor $\phi_{B}$
carry the conserved charge.
In this case, the coupling $\mathit{ABB}$ can be present
and therefore the r.h.s. of inequality~\eqref{eqn:ineq1}
must be replaced by $m_{B}^{2}$ for $\alpha <1$.
Then, $\delta=1/2$ will be an optimal choice, which provides
the analytic strip
$0\leq \mathrm{Im}~q_{1} <  2m_{B}\sqrt{1-1/(4\alpha^{2})}$.
Clearly this strip can be defined only for $\alpha \in (1/2,1]$.}
Solving these inequalities with $\mathrm{Im}\, p_{0} = m_{A}$,
$\vec{p}=\vec{0}$,
and $(q_{0},q_{\perp}) \in \mathbb{R}^{3}$, where
$q_{\perp}= (q_{2},q_{3})$,
we find that the choice 
\bea
\delta = \frac{\alpha^{2}}{2}
\label{eqn:delta}
\eea
provides a maximum analytic strip of Im~$q_{1}$, in which
the vertex function has no singularity:
\bea
0 \leq 
\mathrm{Im}~q_{1} < 
2 m_{B}\sqrt{1-\frac{\alpha^{2}}{4}} = 2 \mu \; .
\label{eqn:imq1strip}
\eea
It is useful to find that 
if $\alpha=1$ ($m_{A}=m_{B} \equiv m$),
the parametrization of external momenta ($\delta=1/2$)
and the analytic strip of Im~$q_{1}$ (maximum value 
is $\sqrt{3}m$) are reduced to the case as discussed
in Ref.~\cite{Luscher:1986dn} for the identical bosonic system.

\par
\begin{figure}[t]
\centering\includegraphics[width=10cm]{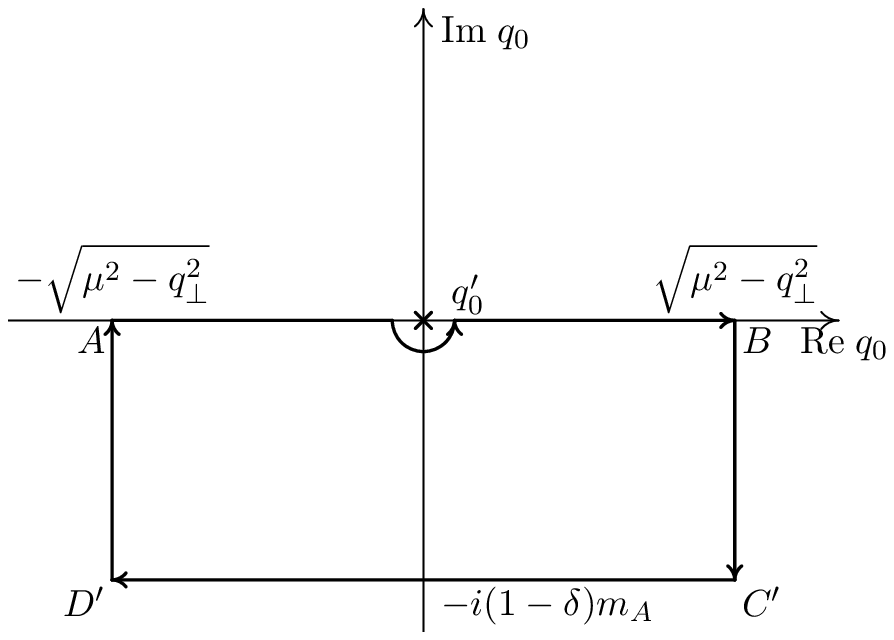}
\caption{$q_{0}$ integral contour for $I_{b1}^{(A)}$.}
\label{fig:q0int1}
\end{figure}
\begin{figure}[t]
\centering\includegraphics[width=10cm]{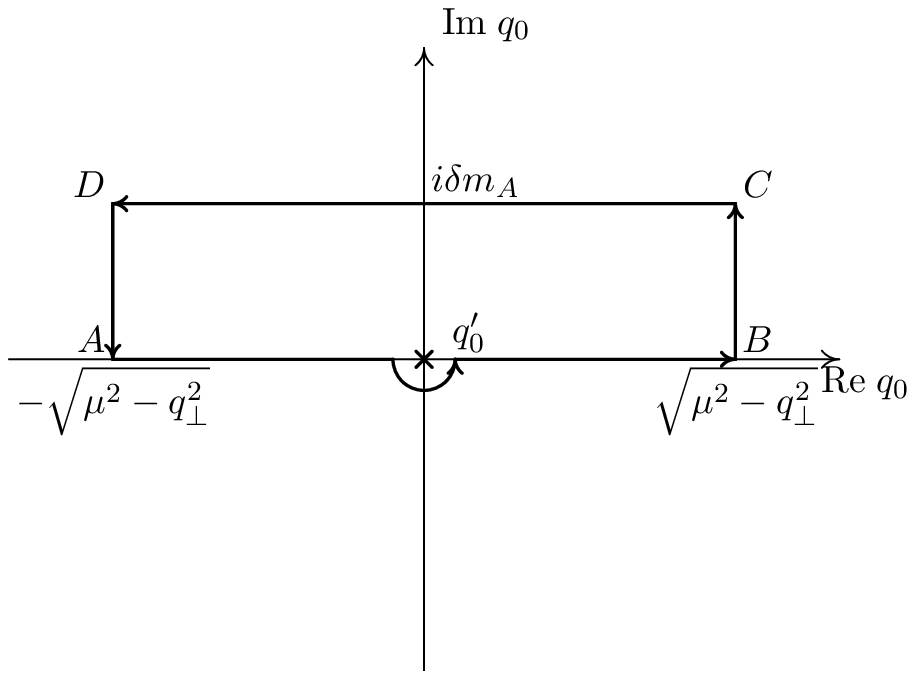}
\caption{$q_{0}$ integral contour for $I_{b1}^{(B)}$.}
\label{fig:q0int2}
\end{figure}

Inserting Eq.~\eqref{eqn:delta}
to the propagators $G_{A}$ and $G_{B}$ in Eq.~\eqref{eqn:integral-b1},
we find that these propagators
possess poles of one-particle states at
\bea
&&
q_{1}^{(A)}=i\sqrt{ q_{0}^{2}+q_{\perp}^{2}+ \mu^{2}
+ i (2-\alpha^{2}) m_{A}q_{0}} \; , \\
&&
q_{1}^{(B)}=i\sqrt{q_{0}^{2}+q_{\perp}^{2}+ \mu^{2}
- i \alpha m_{B} q_{0}} \; ,
\eea
in the complex $q_{1}$ upper half plane, respectively.~\footnote{
For $a$, $b$ $\in \mathbb{R}$, 
\bea
q & \equiv & i \sqrt{a+ib}\nonumber\\*
&=&
-  \sqrt{(\sqrt{a^{2}+b^{2}}-a )/2}
+ i \sqrt{(\sqrt{a^{2}+b^{2}}+a )/2} 
\qquad \mbox{if $b \ge 0$}  \;, \\*
&=& 
+  \sqrt{(\sqrt{a^{2}+b^{2}}-a )/2}
+ i \sqrt{(\sqrt{a^{2}+b^{2}}+a)/2} 
\qquad \mbox{if $b < 0$}  \;.
\eea }
Note that these two poles happen to 
degenerate for $q_{0} = 0$ as
\bea
q_{1}^{(A)} = q_{1}^{(B)} 
= i \sqrt{q_{\perp}^{2}+ \mu^{2}
} 
\; .
\eea
To avoid the double pole in the complex $q_{1}$ plane,
we may deform the $q_{0}$ integration 
path around $q_{0}=0$ into an infinitesimal half circle in the 
complex lower half plane (see Figs~\ref{fig:q0int1} 
and~\ref{fig:q0int2}).
For the purpose of controlling the error term 
associated with integration
up to $O(e^{-L\bar{m}})$ (with $\bar{m}=\sqrt{2}\mu$),
we consider  $q_{0}$ and $q_{\perp}$ in the ball
\bea
\mathbb{B} = 
\{ (q_{0},q_{\perp})\in \mathbb{R}^{3}~|~q_{0}^{2}+q_{\perp}^{2}
\leq 
\mu^{2}
\} \; .
\eea
In this ball it turns out that Im~$q_{1}^{(A)}$ and 
Im~$q_{1}^{(B)}$ are inside the strips
\bea
 m_{B}\sqrt{1-\frac{\alpha^{2}}{4}} 
\leq \mathrm{Im}~q_{1}^{(A)}
\leq 
m_{B}\sqrt{ \frac{1}{\alpha}\sqrt{1-\frac{\alpha^{2}}{4}} 
+1-\frac{\alpha^{2}}{4}}\; ,
\label{eqn:imq1a}
\eea
\bea
 m_{B}\sqrt{1-\frac{\alpha^{2}}{4}} 
\leq \mathrm{Im}~q_{1}^{(B)}
\leq m_{B}\sqrt{\sqrt{1-\frac{\alpha^{2}}{4}}  +1-\frac{\alpha^{2}}{4}} \; ,
\label{eqn:imq1b}
\eea
respectively.  

\par
Now, we choose a contour
depending on  $\alpha$.
If $\alpha \in [\alpha_{c},1]$, where 
$\alpha_{c}\equiv \sqrt{2-\sqrt{2}} \approx 0.765$,
we take a path which goes along the real $q_{1}$ line and
the line Im~$q_{1}=\sqrt{\mu^{2}+m_{A}^{2}}$
closed at $\pm \infty$.
In this case, both $q_{1}^{(A)}$ and $q_{1}^{(B)}$ 
are completely inside the contour.
If $\alpha \in (0,\alpha_{c})$,
we take the line Im~$q_{1}=\sqrt{\mu^{2}+m_{B}^{2}}$ 
instead of $\sqrt{\mu^{2}+m_{A}^{2}}$.
In this case, while $q_{1}^{(B)}$ is completely inside
the contour, $q_{1}^{(A)}$ locates partially outside the contour.
However, since even the lower bound of Im~$q_{1}^{(A)}$
exceeds $\bar{m}$ for $\alpha < \alpha_{c}$
at next step of the complex contour integration of $q_{0}$,
we may neglect the $q_{1}^{(A)}$ contribution here.
In both cases the contribution from the 
path where Im~$q_{1}\ne 0$
is negligible, since it is more rapidly decaying 
at large $L$ than the error term $O(e^{-L\bar{m}})$.
Then, from the residues at $q_{1}^{(A)}$ and $q_{1}^{(B)}$,
we obtain
\bea
I_{b1}
=
I_{b1}^{(A)}+ I_{b1}^{(B)} + O(e^{-L\bar{m}})\; ,
\eea
where
\bea
I_{b1}^{(A)}
&=&
6 i \int_{\mathbb{B}}\frac{dq_{0}d^{2}q_{\perp}}{(2 \pi)^{3}} 
 \frac{e^{i L q_{1}}}{2 q_{1}}
\Gamma_{\mathit{AAB}}(-p,(1-\delta)p+q; \delta p-q)  
G_{B}(\delta p -q)
\nonumber\\*
&&
\times
\Gamma_{\mathit{AAB}}
(p,-(1-\delta) p- q ;  -\delta p + q)  |_{q_{1}=q_{1}^{(A)}} \; ,\\
I_{b1}^{(B)}
&=&
6 i \int_{\mathbb{B}}\frac{dq_{0}d^{2}q_{\perp}}{(2 \pi)^{3}} 
\frac{e^{i L q_{1}}}{2 q_{1}}
 \Gamma_{\mathit{AAB}}(-p, (1-\delta)p+q;  \delta p -q)  
 \nonumber\\*
&&
\times G_{A}((1-\delta)p+q)
\Gamma_{\mathit{AAB}}(p,-(1-\delta )p-q; - \delta p +q) 
|_{q_{1}=q_{1}^{(B)}} \; .
\eea
As mentioned above, 
$I_{b1}^{(A)}$ is  relevant only
for the case $\alpha \in  [\alpha_{c},1]$.
Note that 
$G_{B}( \delta p -q)|_{q_{1}=q_{1}^{(A)}}$
and
$G_{A}((1-\delta)p+q)|_{q_{1}=q_{1}^{(B)}}$
have a common pole at $q_{0}=0$.

\par
Secondly, we perform the complex contour integration of $q_{0}$.
$I_{b1}^{(A)}$ is evaluated along the contour in Fig.~\ref{fig:q0int1}.
The contributions from the paths $\mathit{BC'}$ and $\mathit{D'A}$ 
are negligible at large $L$, since Im~$q_{1}^{(A)}\geq \bar{m}$ 
along these lines.
The choice of the contour $\mathit{C'D'}$ 
(given by the shift $q_{0} \to q_{0} - i (1-\delta)m_{A}$) is 
possible because 
both inequalities \eqref{eqn:ineq1} and \eqref{eqn:ineq2}
are satisfied as long as $\alpha > \sqrt{3-\sqrt{7}} \approx 0.595$
and thus the vertex function has no singularity inside the 
contour for $\alpha \in [\alpha_{c},1]$.
As there is no pole inside the contour,
the integral can be written as that along the path 
$\mathit{C'D'}$.
The argument of the propagator $G_{B}$ and $q_{1}^{(A)}$ are 
then shifted as $\delta p-q \to p-q$ and 
$q_{1}^{(A)} \to i \sqrt{q_{0}^{2}+q_{\perp}^{2}+m_{A}^{2}}$, 
respectively, where
$q$ satisfies the on-shell condition $q^{2} = -m_{A}^{2}$.
The integral is then represented as
\bea
I_{b1}^{(A)}
&= &
6 \int_{\mathbb{B}}\frac{dq_{0}d^{2}q_{\perp}}{(2 \pi)^{3}}
\frac{e^{-L\sqrt{q_{0}^{2}+q_{\perp}^{2}+m_{A}^{2}}}}
{2\sqrt{q_{0}^{2}+q_{\perp}^{2}+m_{A}^{2}}}
\; 
F_{\mathit{AA}}^{(b1-2)}(iq_{0}) 
+O(e^{-L\bar{m}})
\; ,
\label{eqn:ib1p}
\eea
where
\bea
F_{\mathit{AA}}^{(b1-2)}(iq_{0}) 
&=& 
\Gamma_{\mathit{AAB}}(-p, q; p-q)  G_{B}(p -q) 
\nonumber\\*
&&
\times
\Gamma_{\mathit{AAB}}(p, -q; -p+q)
|_{p^{2}=q^{2}=-m_{A}^{2}}
\eea
is a one-particle-irreducible (1PI) part of the forward scattering 
amplitude of $F_{\mathit{AA}}(\nu)$
with $\nu=pq/m_{A}=iq_{0}$
(see the diagram (b1-2) in Fig.~\ref{fig:aa-scattering}).

\par
$I_{b1}^{(B)}$ is then evaluated 
along the contour in Fig.~\ref{fig:q0int2}.
Again the contributions from the paths $\mathit{BC}$
and $\mathit{DA}$ are negligible at large $L$, since
Im~$q_{1}^{(B)}\geq \bar{m}$ along these lines.
The integral along the path $\mathit{CD}$ is parametrized 
by shifting the momentum as $q_{0} \to q_{0} + i \delta m_{A}$,
which is possible because the vertex function is
analytic inside the contour for the whole range of $\alpha \in (0,1]$.
The argument of $G_{A}$ and $q_{1}^{(B)}$ are shifted as
$(1-\delta) p+q \to p+q$ and $q_{1}^{(B)} \to 
i\sqrt{q_{0}^{2}+q_{\perp}^{2}+m_{B}^{2}}$ and then
$q^{2}=-m_{B}^{2}$.
Now, since the pole at $q_{0}=0$ is
inside the contour, we also have a residue contribution.
Thus the integral becomes
\bea
I_{b1}^{(B)}
&= &
3  
\int_{\mathbb{B}'} \frac{d^{2}q_{\perp}}{(2 \pi)^{2}}
\frac{e^{- L\sqrt{ q_{\perp}^{2} + \mu^{2} }
}}{2\sqrt{ q_{\perp}^{2}+\mu^{2}}} \;
\frac{\lambda_{AAB}^{2}}{2\nu_{B}}
\nonumber\\*
&&
+ \; 
 6  \int_{\mathbb{B}} \frac{dq_{0}d^{2}q_{\perp}}{(2 \pi)^{3}}
\frac{e^{-L\sqrt{q_{0}^{2}+q_{\perp}^{2}+m_{B}^{2}}}}
{2\sqrt{q_{0}^{2}+q_{\perp}^{2}+m_{B}^{2}}}
\;  F_{\mathit{AB}}^{(b1-1)}(iq_{0}) 
+O(e^{-L\bar{m}}) \; ,
\label{eqn:ib1m}
\eea
where
\bea
F_{\mathit{AB}}^{(b1-1)}(iq_{0}) 
&=&
\Gamma_{\mathit{AAB}}(-p, p+q; -q)  G_{A}(p+q)
 \nonumber\\*
 &&
 \times
\Gamma_{\mathit{AAB}}(p, - p-q ; q) 
|_{p^{2}=-m_{A}^{2},\; q^{2}=-m_{B}^{2}}
\eea
is a 1PI part of the forward scattering amplitudes
of $F_{\mathit{AB}}(\nu)$
(see the diagram (b1-1) in Fig.~\ref{fig:ab-scattering}).
Note that $F_{\mathit{AB}}^{(b1-1)}(\nu)$ has a pole at $\nu=+\nu_{B}$.
The coupling $\lambda_{\mathit{AAB}}$ in the first 
term in Eq.~\eqref{eqn:ib1m}
is the effective renormalized coupling 
defined through the relation
\bea
\frac{\lambda_{\mathit{AAB}}^{2}}{2}  
&=&
\lim_{\nu \to \pm \nu_{B}}(\nu^{2}-\nu_{B}^{2})F_{\mathit{AB}}(\nu)
=
\lim_{\nu \to +\nu_{B}}(\nu^{2}-\nu_{B}^{2})F_{\mathit{AB}}^{(b1-1)}(\nu)
\nonumber\\
&=&
\frac{\nu_{B}}{m_{A}}  \Gamma_{\mathit{AAB}}(-p, p+q; -q)  
\Gamma_{\mathit{AAB}}(p, - p-q ; q) |_{\nu = +\nu_{B}} \; ,
\label{eqn:effcoupling-s}
\eea
where all legs of the vertex function is on the mass shell, 
$p^{2}=-m_{A}^{2}$, $(p + q)^{2}=-m_{A}^{2}$ and $q^{2}=-m_{B}^{2}$.
The integral region of $q_{\perp}$ in the first term 
in Eq.~\eqref{eqn:ib1m} is defined by
\bea
\mathbb{B}' = \{ q_{\perp} \in \mathbb{R}^{2}
~|~ q_{\perp}^{2} \leq   \mu^{2}  \} \; .
\eea

\par
One may find that the 1PI amplitudes corresponding to (b1-1) in 
Fig.~\ref{fig:aa-scattering} and (b1-2) in Fig.~\ref{fig:ab-scattering},
which are necessary to build up 
the forward scattering amplitudes in the mass shift formula,
are apparently absent in $I_{b1}$.
However, we find that these amplitudes can 
be included by using the crossing relation 
$F_{\mathit{AA}}^{(b1-1)}(-\nu)=F_{\mathit{AA}}^{(b1-2)}(\nu)$
and
$F_{\mathit{AB}}^{(b1-1)}(-\nu)=F_{\mathit{AB}}^{(b1-2)}(\nu)$; 
one can replace 
$F_{\mathit{AA}}^{(b1-2)}(\nu)$ 
in Eq.~\eqref{eqn:ib1p} by 
$(F_{\mathit{AA}}^{(b1-1)}(\nu)+F_{\mathit{AA}}^{(b1-2)}(\nu))/2$
and 
$F_{\mathit{AB}}^{(b1-1)}(\nu)$ 
in Eq.~\eqref{eqn:ib1m} by
$(F_{\mathit{AB}}^{(b1-1)}(\nu)+F_{\mathit{AB}}^{(b1-2)}(\nu))/2$.

\par
Finally, we carry out the $q_{\perp}$ integration 
in Eqs.~\eqref{eqn:ib1p} and~\eqref{eqn:ib1m}
using the formula
\bea
\int_{-\infty}^{\infty} \frac{d^{2}q_{\perp}}{(2 \pi)^{2}}
\frac{e^{- L\sqrt{q_{\perp}^{2}+ \rho^{2}}}}
{2 \sqrt{q_{\perp}^{2}+\rho^{2}}}
=
\frac{1}{4\pi L}e^{- L \rho} \; ,
\label{eqn:formula_qperp_int}
\eea
where  the integral region
is extended from $\mathbb{B}$ or $\mathbb{B}'$ to the infinity, 
because the additional contributions are always
up to the order of the error term.
Hence, we end up with
\bea
I_{b1}
&= &
\frac{3  }{4 \pi L}
\Biggl [
\frac{\lambda_{\mathit{AAB}}^{2}}{2 \nu_{B}} e^{-L  \mu }
\nonumber\\*
&&
+  \int_{-\infty }^{\infty}  \! \! \frac{dq_{0}}{2 \pi} \; 
e^{-L\sqrt{q_{0}^{2}+m_{A}^{2}}}
\; 
\{ F_{\mathit{AA}}^{(b1-1)}(iq_{0})+F_{\mathit{AA}}^{(b1-2)}(iq_{0}) \}
\nonumber\\*
&& 
+
\int_{-\infty}^{\infty}\! \!  \frac{dq_{0}}{2 \pi}\;
e^{-L\sqrt{q_{0}^{2}+m_{B}^{2}}} \; 
\{ F_{\mathit{AB}}^{(b1-1)}(iq_{0})+F_{\mathit{AB}}^{(b1-2)}(iq_{0}) \}
\Biggr ]  
\! \!  +O(e^{-L\bar{m}})
\; .
\label{eqn:boson-result}
\eea
By combining the contributions from the other self-energy diagrams
in Fig.~\ref{fig:selfenergy} (see Appendix~\ref{sec:summary-result}), 
we complete Eq.~\eqref{eqn:massshift-boson}.

\subsection{Fermion mass shift formula}

Let us next consider the fermion mass shift formula in 
the $\Psi_{A}$-$\phi_{B}$ system.
The self-energy diagram~(b1) in 
Fig.~\ref{fig:selfenergy-fermion} with $|\vec{m}|=1$,
sandwiched by the spinors $\bar{u}(p,r)$ and $u(p,r)$, is
given by
\bea
I_{b1} &=& 
\int \frac{d^{4}q}{(2 \pi)^{4}}\;
\left ( 2 \sum_{j=1}^{3} \cos L q_{j}\right )
\; 
\bar{u}(p,r)
\Gamma_{\mathit{AAB}}(-p, (1-\delta)p+q ;  \delta p-q) 
 \nonumber\\*
&& 
\times 
S_{A}( (1-\delta)p+q) G_{B}( \delta p - q) 
\Gamma_{\mathit{AAB}}
(p, -(1-\delta)p-q ; -\delta p + q)  
\nonumber\\*
&& 
\times
u(p,r)
\biggr |_{p=(im_{A}, \; \vec{0} \, )}\; .
\label{eqn:integral-b1-fermion}
\eea 
Since the pole structure of the integrand is the same 
as in the $\phi_{A}$-$\phi_{B}$ system,
we can carry out the momentum integration
in a similar way.

\par
After several integration steps, we then arrive at
the expressions 
(see Figs.~\ref{fig:aa-fermion-scattering} 
and~\ref{fig:ab-scattering}, respectively):
\bea
&&
I_{b1}^{(A)} 
= 
- 3\!
\int \!\frac{dq_{0}d^{2}q_{\perp}}{(2 \pi)^{3}}
\frac{e^{-L\sqrt{q_{0}^{2}+q_{\perp}^{2}+m_{A}^{2}}}}
{2\sqrt{q_{0}^{2}+q_{\perp}^{2}+m_{A}^{2}}} \; 
\!
\{ 
F_{\mathit{AA}}^{(b1)}(iq_{0}) + 
F_{A\bar{A}}^{(b1)}(iq_{0})
\} 
\! +O(e^{-L\bar{m}})
\; ,\nonumber\\
&&\\*
&&
F_{\mathit{AA}}^{(b1)}(iq_{0}) 
=
-
\sum_{s} \bar{u}_{\alpha}(p,r)
\Gamma_{\mathit{AAB}}^{\alpha\beta}
(-p, q; p- q) u_{\beta}(q,s)   G_{B} (p -q) \nonumber\\*
&& 
\quad\quad\quad\quad\quad\quad
\times 
\bar{u}_{\gamma}(q,s) 
\Gamma_{\mathit{AAB}}^{\gamma\delta}
(p, -q ; -p+ q)u_{\delta} (p,r) 
|_{p^{2}=q^{2}=-m_{A}^{2}} \; ,
\label{eqn:fermion-aa-scattering}\\*
&&
F_{A\bar{A}}^{(b1)}(iq_{0})
=
\sum_{s}
\bar{u}_{\alpha}(p,r)\Gamma_{\mathit{AAB}}^{\alpha\beta}
(-p,-q; p+q ) v_{\beta}(q,s) G_{B} (p + q )\nonumber\\*
&&
\quad\quad\quad\quad\quad\quad
\times 
\bar{v}_{\gamma}(q,s) \Gamma_{\mathit{AAB}}^{\gamma\delta}
(p,q; -p-q) 
u_{\delta} (p,r) |_{p^{2}=q^{2}=-m_{A}^{2}} \; ,
\label{eqn:fermion-aabar-scattering}
\eea
and
\bea
&&
I_{b1}^{(B)}
=
3  \int  \frac{d^{2}q_{\perp}}{(2 \pi)^{2}}
\frac{e^{- L
\sqrt{ q_{\perp}^{2} + \mu^{2}}
}}{2\sqrt{ q_{\perp}^{2} +\mu^{2}}} \;
\frac{\lambda_{AAB}^{2}}{2\nu_{B}}
 \nonumber\\*
 &&
\quad
 + \;
3 \int_{}^{}\frac{dq_{0}d^{2}q_{\perp}}{(2 \pi)^{3}}
\frac{e^{-L\sqrt{q_{0}^{2}+q_{\perp}^{2}+m_{B}^{2}}}}
{2\sqrt{q_{0}^{2}+q_{\perp}^{2}+m_{B}^{2}}}
\{ F_{\mathit{AB}}^{(b1-1)}(iq_{0})  
+F_{\mathit{AB}}^{(b1-2)}(iq_{0})   \}
+O(e^{-L\bar{m}})
\; , 
\nonumber\\*
&&\\*
&&
F_{\mathit{AB}}^{(b1-1)}(iq_{0}) 
= 
\bar{u}_{\alpha}(p,r)
\Gamma_{\mathit{AAB}}^{\alpha\beta}(-p, p+q ; -q) 
S_{A}^{\beta\gamma}(p + q ) 
\nonumber\\*
&&
\quad\quad\quad\quad\quad\quad
\times
\Gamma_{\mathit{AAB}}^{\gamma\delta}(p, -p-q; q) 
u_{\delta} (p,r) 
|_{p^{2}=-m_{A}^{2}, \; q^{2}=-m_{B}^{2}} \; ,\\
&&
F_{\mathit{AB}}^{(b1-2)}(iq_{0})
= 
\bar{u}_{\alpha}(p,r)\Gamma_{\mathit{AAB}}^{\alpha\beta}
(-p,p-q; q) S_{A}^{\beta\gamma}(p - q )  \nonumber\\*
&& 
\quad\quad\quad\quad\quad\quad
\times
\Gamma_{\mathit{AAB}}^{\gamma\delta}
(p,-p+q; -q) u_{\delta} (p,r) 
|_{p^{2}=-m_{A}^{2}, \; q^{2}=-m_{B}^{2}} \; ,
\eea
where we have used the relations
$m_{A}-i\sla{q} = \sum_{s} u(q,s)\bar{u}(q,s)$
and $m_{A}+i\sla{q} = -\sum_{s} v(q,s)\bar{v}(q,s)$
in Eqs.~\eqref{eqn:fermion-aa-scattering} 
and~\eqref{eqn:fermion-aabar-scattering}, respectively.
Repeated Greek letters, corresponding to spinor components,
are implicitly summed.
The overall minus sign in $F_{\mathit{AA}}^{(b1)}$
is due to Fermi statistics.
The effective coupling $\lambda_{\mathit{AAB}}$
is defined as in Eq.~\eqref{eqn:effcoupling-s}.
The advantage of such a definition 
of the coupling 
in the fermion-boson system
is that the model dependent matrix structure of the
vertex function, such as 
$\{I,\gamma_{\mu},\sigma_{\mu\nu},\gamma_{5}\gamma_{\mu},
\gamma_{5} \}$, is involved in $\lambda_{\mathit{AAB}}$
through the definition of $F_{\mathit{AB}}$.

\par
After integrating out $q_{\perp}$, 
we obtain the expression
\bea
I_{b1}
&=&
\frac{3}{4 \pi L}
\Biggl [
\frac{ \lambda_{\mathit{AAB}}^{2}}{2 \nu_{B}}
e^{-L\mu }
\nonumber\\*
&&
- \! \int_{-\infty }^{\infty}  \! \frac{dq_{0}}{2 \pi} \; 
e^{-L\sqrt{q_{0}^{2}+m_{A}^{2}}}
\{ F_{\mathit{AA}}^{(b1)}(iq_{0}) 
+ F_{\mathit{A\bar{A}}}^{(b1)}(iq_{0}) \}
\nonumber\\*
&&
+  \! \int_{-\infty}^{\infty} \!
\frac{dq_{0}}{2 \pi}  \;
e^{-L\sqrt{q_{0}^{2}+m_{B}^{2}}}
\{ F_{\mathit{AB}}^{(b1-1)}(iq_{0}) +F_{\mathit{AB}}^{(b1-2)}(iq_{0}) \}
\Biggr ] \! +O(e^{-L\bar{m}})\; .
\label{eqn:fermion-result}
\eea
We find that 
the expression is almost 
equivalent to Eq.~\eqref{eqn:boson-result}
apart from the relative minus sign in the second term.
By combining the contribution from the other self-energy diagrams  
(see Appendix~\ref{sec:summary-result}),
we complete Eq.~\eqref{eqn:massshift-fermion}.

\section{The nucleon mass shift}
\label{sec:nucleon}

As an application of the fermion mass shift 
formula in Eq.~\eqref{eqn:massshift-fermion},
let us discuss the nucleon mass shift in the $N$-$\pi$ system.
Since the formula is expected to hold 
nonperturbatively,
it is interesting to estimate the 
mass shift by inserting the $N$-$\pi$ scattering
amplitude which is known from experiment.
Therefore, the following analysis
can be regarded as an estimate of the finite volume
effect on the nucleon mass 
when the realistic pion mass is achieved in lattice
QCD simulations with dynamical fermions.

\par
According to H\"ohler~\cite{Hohler:1984ux}, the 
subthreshold expansion of the $N$-$\pi$ forward scattering 
amplitude around $\nu = 0$ is parametrized as
\bea
F_{\mathit{N\pi}}(\nu) = 6 m_{N} D^{+}(\nu)\; ,
\label{eqn:n-pi-famp}
\eea
where
\bea
D^{+} (\nu)
= 
\frac{g^{2}}{m_{N}}\frac{\nu_{B}^{2}}{\nu_{B}^{2}-\nu^{2}}
+d_{00}^{+} \; m_{\pi}^{-1} +d_{10}^{+} \; m_{\pi}^{-3} \nu^{2}
+d_{20}^{+} \; m_{\pi}^{-5} \nu^{4} + O(\nu^{6}) \; .
\label{eqn:d-plus}
\eea
The isospin sum is taken in Eq.~\eqref{eqn:n-pi-famp} and
the effect of isospin symmetry breaking is neglected.
The masses of the nucleon and pion
are then given by $m_{N}=938$ MeV and $m_{\pi}=140$ MeV.
The coupling constant is $g^{2}/4 \pi =14.3$.
The first term in Eq.~\eqref{eqn:d-plus}
is identified with the pseudovector nucleon Born term
with $\nu_{B}=m_{\pi}^{2}/2m_{N} \approx 0.07 m_{\pi}$.
The coefficients of the other terms are given by 
$d_{00}^{+} = -1.46(10)$, $d_{10}^{+}=1.12(2)$ and 
$d_{20}^{+} = 0.200(5)$~\cite{Hohler:1984ux}.
We only take into account the mean of these values hereafter.
Note that although Eq.~\eqref{eqn:n-pi-famp} 
is defined in Minkowski space-time such that 
$\nu$ is a real variable here, 
it can directly be inserted to 
Eq.~\eqref{eqn:massshift-fermion} by 
rotating this variable as 
$\nu=iq_{0}$, where
$q_{0}$ is the integral variable in 
Eq.~\eqref{eqn:massshift-fermion}.
The effective coupling is computed by
using Eq.~\eqref{eqn:eff-re-coupling} as
\bea
\lim_{\nu \to \pm \nu_{B}}(\nu^{2}-\nu_{B}^{2})
F_{\mathit{N\pi}}(\nu)=
\frac{\lambda_{\mathit{NN\pi}}^{2}}{2}
=  - 6 g^{2} \nu_{B}^{2} \; .
\eea
Since  $m_{\pi}/m_{N}=0.149 <\alpha_{c}$, we may 
neglect the second term in
Eq.~\eqref{eqn:massshift-fermion}.
It is clear that since the factor of the pole term 
in Eq.~\eqref{eqn:massshift-fermion} is already
twice larger than that in Eq.~\eqref{eqn:formula-doubt},
we obtain a different expression from that in 
Ref.~\cite{Luscher:1983rk}.
In fact, the mass shift formula, divided by 
the nucleon mass itself, is reduced to
\bea
\delta (\xi \! = \! Lm_{\pi}) 
&\equiv&  \Delta m_{N}(L)/m_{N} 
\nonumber\\*
&=&
\frac{9}{2 \xi} \left (\frac{g^{2}}{4 \pi} \right )
\left (\frac{m_{\pi}}{m_{N}} \right )^{3} 
e^{- \xi \sqrt{1- \nu_{B}^{2}/m_{\pi}^{2}}}
\nonumber\\*
&&
- \frac{3}{16 \pi^{2} \xi} 
\left (\frac{m_{\pi}}{m_{N}} \right )^{2} \!\!
\int_{-\infty}^{\infty}dy \;
e^{-\xi \sqrt{1+y^{2}}} F_{\mathit{N\pi}} (i m_{\pi} y) 
+O(e^{-L\bar{m}})
\nonumber\\*
&=&
\delta_{P}(\xi)+\delta_{B}(\xi)+\delta_{R} (\xi)+O(e^{-L\bar{m}})\; ,
\label{eqn:nucleon-massshift}
\eea
where $\delta_{P}(\xi)$ is the pole term,
and $\delta_{B}(\xi)$ and $\delta_{R}(\xi)$ 
correspond to the contributions of the pseudovector 
Born term and the rest in Eq.~\eqref{eqn:d-plus}.
Clearly, this formula recovers the 
expression given in the framework of ChPT~\cite{AliKhan:2003cu}.

\begin{figure}[!t]
\centering\includegraphics[width=11cm]{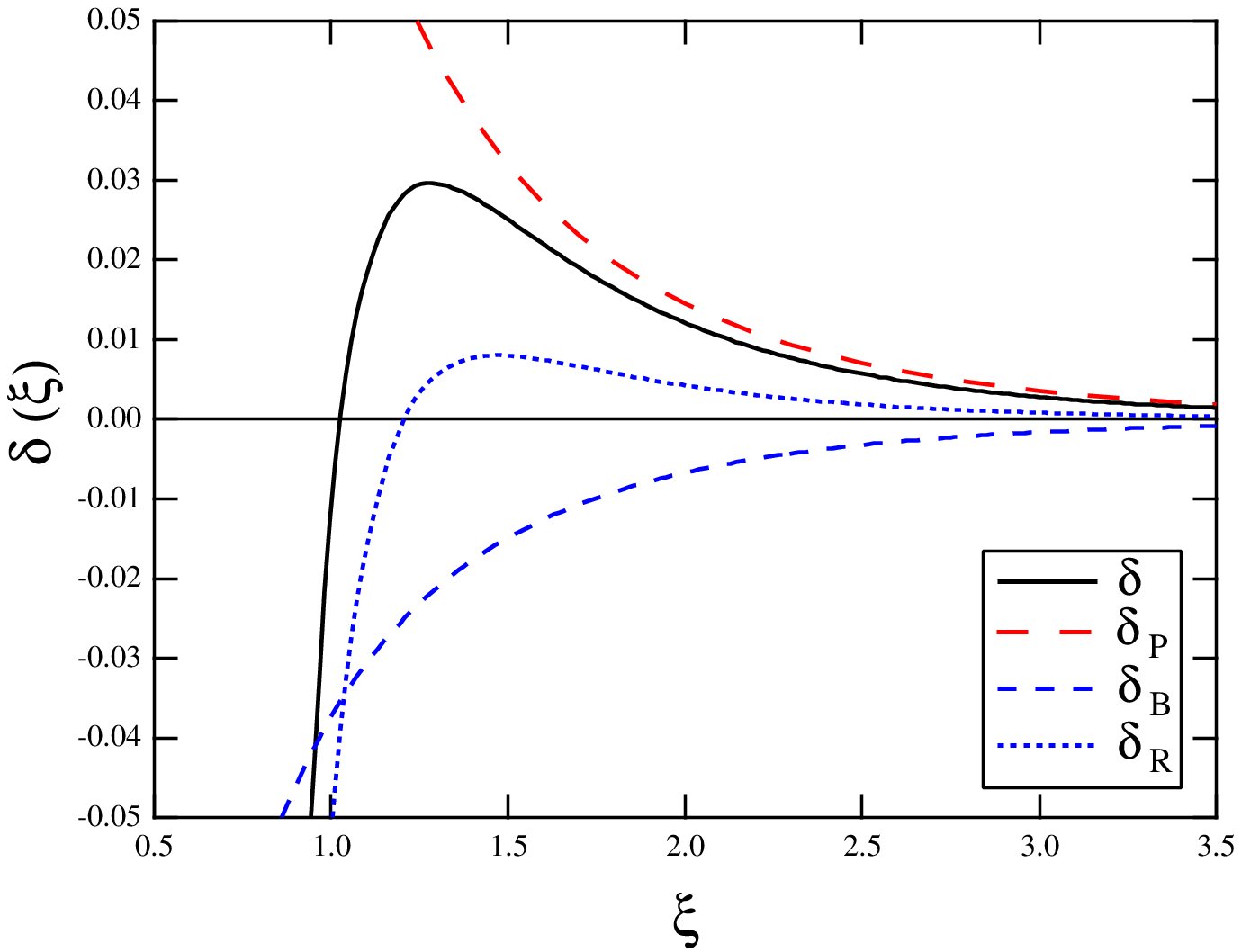}
\caption{The nucleon mass shift as a function of $\xi=L m_{\pi}$.} 
\label{fig:delta_xi}
\end{figure}

\par
We  plot $\delta (\xi)$ in Fig.~\ref{fig:delta_xi},
where $\xi =1$ corresponds to $L=1.4$ fm.
We find that $\delta (\xi)$ suffers strongly from higher 
order contributions of $\nu$ in the range $\xi \leq 1$.
For instance, $\delta_{R} (\xi)$ causes the negative mass 
shift within the leading mass shift formula ($|\vec{m}|=1$).
In this range, the contribution from $|\vec{m}| \geq \sqrt{2}$ 
to the formula, of course, will not be negligible.
On the other hand, $\delta (\xi)$ seems to be 
mostly described by $\delta_{P}(\xi)$ as $\xi$ increases.
However we notice that this is due to the cancellation between
$\delta_{B}(\xi)$ and $\delta_{R}(\xi)$.

\par
Finally, let us discuss the relation between the 
general fermion mass shift formula
in Eq.~\eqref{eqn:massshift-fermion} and 
the nucleon mass shift formula derived in Ref.~\cite{AliKhan:2003cu}.
The diagrams that they evaluated within ChPT
correspond to (b1) and (a2) in Fig.~\ref{fig:selfenergy-fermion},
where the vertex functions were treated as certain
coupling constants.
The diagram~(b1) was evaluated
by introducing the Feynman parameter $x$, where
its initial integral range $x \in [0,1]$ was modified as
$[0,1]=(-\infty,\infty)- (-\infty,0)- (1,\infty)$ 
and the terms of $O(e^{-Lm_{N}})$ which mainly originate
from the integral in the interval $(1,\infty)$ 
were omitted according to 
the infrared regularization~\cite{Becher:1999he}.
Then, from the integrals in the range
$(-\infty,\infty)$ and $(-\infty,0)$,
they obtained the terms 
corresponding to
$\delta_{P}(\xi)$ and $\delta_{B}(\xi)$
in the mass shift formula, respectively.
The evaluation of the diagram (a2) was straightforward
and they obtained terms 
for $\delta_{R}(\xi)$.
In our point of view this is reasonable
since the diagram (a2) has nothing to do with the pole term.
Here, it is interesting to notice that the omitted
terms according to the infrared regularization
can exactly be used  to reconstruct
the terms involving $F_{NN}(\nu)$ and $F_{N\bar{N}}(\nu)$,
which we have also omitted in Eq.~\eqref{eqn:nucleon-massshift} 
because of $\alpha <\alpha_{c}$.
In this sense, the general formula 
in Eq.~\eqref{eqn:massshift-fermion} covers
all possible contributions to the mass shift,
some of which may be neglected depending on the
physical situation.

\section{Summary}
\label{sec:summary}

We have studied the general finite size mass shift formula 
for the two stable distinguishable particle system
in a periodic finite volume along the lines of L\"uscher's proof
for an identical bosonic theory.
The main results are Eqs.~\eqref{eqn:massshift-boson}
and~\eqref{eqn:massshift-fermion}.
What we have clarified are the following.

\par
The boson mass shift formula for $\phi_{A}$ 
in the $\phi_{A}$-$\phi_{B}$ system is related to
{\em two} types of the forward scattering amplitudes, 
$F_{\mathit{AB}}(\nu)$ and $F_{\mathit{AA}}(\nu)$.
If $m_{B}/m_{A} < \alpha_{c} =\sqrt{2-\sqrt{2}} \approx 
0.765$, the contribution from 
the $F_{\mathit{AA}}(\nu)$ term 
becomes smaller than the order of the error term.
It has turned out that
the speculation such as in Eq.~\eqref{eqn:formula-doubt}
is not valid for a two particle system.
However, the mechanism how the mass shift, proportional to
the difference of the self energy $\Sigma_{L}-\Sigma$,
is related to the forward scattering amplitude is
the same as in the identical bosonic system~\cite{Luscher:1986dn}.
In principle, the difference of the self energy
consists  of the propagators of $\phi_{A}$ and $\phi_{B}$,
one of which is broken into external legs on the mass shell
after integration over the loop momentum.
Then, a self-energy diagram is related to one (or some) of 
the 1PI parts of the forward scattering amplitude.
It is to be noted that in the two particle system,
unless the crossing symmetry among the 1PI parts of the 
amplitude is made manifest, some of the 1PI parts necessary 
to build up the forward scattering
amplitudes will apparently be missed although 
such an expression is mathematically consistent.
(In the identical particle system, 
the crossing symmetry is automatically manifest.)

\par
The fermion mass shift formula for $\Psi_{A}$ 
in the $\Psi_{A}$-$\phi_{B}$ system is related to
{\em three} types of the forward scattering amplitudes, 
$F_{\mathit{AB}}(\nu)$, $F_{\mathit{AA}}(\nu)$ 
and $F_{\mathit{A\bar{A}}}(\nu)$.
If $m_{B}/m_{A} < \alpha_{c}$, the contributions from the 
$F_{\mathit{AA}}(\nu)$ 
and $F_{\mathit{A\bar{A}}}(\nu)$ terms 
become smaller than the order of the error term.
Again,  crossing symmetry must be taken into account
to relate the self-energy diagrams to
all 1PI parts of the forward scattering amplitudes.
When the fermion propagator is
broken into external legs on the mass shell,
the relative minus sign appears in front of
the corresponding terms.

\par
In both cases,  $\lambda_{\mathit{AAB}} = 0$
unless the self energy contains the (b1) diagram,
since it means that $F_{\mathit{AB}}(\nu)$ has no pole.

\par
Finally, we have written down the nucleon mass shift formula
by applying the fermion mass shift formula in 
Eq.~\eqref{eqn:massshift-fermion} to the $N$-$\pi$ system, 
thereby we have found that the pole term is
underestimated by factor two in L\"uscher's formula in 
Ref.~\cite{Luscher:1983rk} and 
the formula derived in ChPT~\cite{AliKhan:2003cu} 
is reproduced.

\par
There are now general
finite size mass shift formulae 
for the two particle systems which are valid to 
all orders in perturbation theory within $|\vec{m}| =1$, 
in addition to that in the identical bosonic 
system~\cite{Luscher:1986dn}.
For every case all these formulae are obtained in the same way
by carefully analyzing the appropriate set of self-energy diagrams.

\section*{Acknowledgments}

We are grateful to P.~Weisz for introducing us to this interesting 
topic and also for numerous discussions during the course of the 
present work.
We also appreciate useful comments from M.~L\"uscher
and G.~Colangelo.
We are partially supported  by the 
DFG Forschergruppe `Lattice Hadron Phenomenology.'
M.K. is also supported by Alexander von Humboldt
foundation, Germany.

\appendix
\section{Notation}
\label{sec:notation}

In this appendix, we summarize our notations.

\subsection{Boson propagator}

The boson propagator in infinite volume in Euclidean space 
is defined by
\bea
&&
G (x-y)  =
\int \frac{d^{4}p}{(2 \pi)^{4}} G(p) e^{ip(x-y)} \; ,\\
&& 
G(p) = \frac{1}{p^{2}+m^{2} - \Sigma(p)} \; ,\\
&&
\Sigma(p)=\frac{\partial}{\partial p^{2}}\Sigma(p)=0
\quad \mbox{for} \quad p^{2}=-m^{2} \;  ,
\label{eqn:renormalization-boson}
\eea
where $m$ is the physical mass.
The renormalization conditions~\eqref{eqn:renormalization-boson}
mean that the propagator has pole 
at $p^{2} = - m^{2}$ with unit residue.
We assume that there are no bound states so that 
this pole is the only singularity of
$G(p)$ below the two-particle threshold.

\par
In the finite volume of the size $L^{3}$, by
imposing periodic boundary conditions for the spatial
directions,  we have 
\bea
&&
 G_{L} (x-y)  =
\frac{1}{L^{3}}\sum_{\vec{p}}
\int \frac{dp_{0}}{2 \pi}
G_{L}(p) e^{ip(x-y)} \; ,\\
&&
G_{L}(p) = \frac{1}{p^{2}+m^{2} -\Sigma_{L}(p)} \; ,
\eea
where the momenta are quantized as
$\vec{p}=2 \pi \vec{n}/L$ ($\vec{n} \in \mathbb{Z}^{3}$).
The summation of discrete momenta 
can be reformulated as the integral by using the Poisson
summation formula (see Eq.~\eqref{eqn:poisson}).
We expect that the pole position of $G_{L}(p)$ 
is shifted from $p^{2}=-m^{2}$ to $p^{2}=-M(L)^{2}$ 
with $M(L) \equiv m+\Delta m(L)$.

\par
In the $\phi_{A}$-$\phi_{B}$ system, 
by expanding $\Sigma_{L}(p)$ in powers of $\Delta m_{A} (L)$,
we then solve
\bea
G_{L}(p)^{-1} |_{p^{2}=-M_{A}(L)^{2}}=0 \; ,
\eea
and obtain Eq.~\eqref{eqn:massshift-2boson}
for $\Delta m_{A}(L)$.

\subsection{Fermion propagator}

The fermion propagator in infinite volume 
in Euclidean space is
\bea
&&
S (x-y)  =
\int \frac{d^{4}p}{(2 \pi)^{4}}
S(p) e^{ip(x-y)} \; ,\\
&& 
S(p) = \frac{1}{i\sla{p}+m -\Sigma(p) } \; ,\\
&&
\Sigma(p)=\frac{\partial}{\partial \sla{p}}\Sigma(p)=0
\quad \mbox{for} \quad  i \sla{p} =  -m  \; ,
\eea
where $\sla{p} \equiv \gamma_{0}^{E}p_{0}+\gamma_{i}^{E} p_{i}$.
Euclidean Dirac matrices are defined by
$\gamma_{0}^{E}=\gamma_{0}$ and
$\gamma_{i}^{E}= i \gamma_{i}=-i\gamma^{i}$ ($i=1,2,3$),
which satisfy
$\{ \gamma_{\mu}^{E},\gamma_{\nu}^{E} \} = 2 \delta_{\mu\nu}$.
As similar to the boson propagator, we also assume that there 
are no bound states so that the pole at
$i\sla{p}=-m$ is the only singularity of
$S(p)$ below the two-particle threshold.

\par
In the finite volume of the size $L^{3}$
with periodic boundary conditions, we have 
\bea
&&
S_{L} (x-y)  =
\frac{1}{L^{3}}\sum_{\vec{p}}
\int \frac{dp_{0}}{2 \pi}
S_{L}(p) e^{ip(x-y)} \; ,\\
&&
S_{L}(p) = \frac{1}{i\sla{p}+m-\Sigma_{L}(p)} \; ,
\eea
where we expect that the pole position is shifted from
$i\sla{p}=-m$ to $i\sla{p} = - M(L)$ 
with $M(L)=m+\Delta m(L)$.

\par
In the $\Psi_{A}$-$\phi_{B}$ system, 
we then solve the equation
\bea
\bar{u}(p,r) S_{L}(p)^{-1} u(p,r) |_{p^{2}=-M_{A}(L)^{2}} = 0\; ,
\eea
where $\bar{u}(p,r)$ and  $u(p,r)$ are 
spinors of $\Psi_{A}$ defined as the 
solution of the free Dirac equation
\bea
&&
(i\sla{p} + m) u (p,r)=0 \; , \quad 
\bar{u} (p,r)(i\sla{p} +m) =0 \; , \nonumber\\
\nonumber\\*
 &&
\bar{u} (p,r) u(p,s) = 2 m \delta_{rs}\; ,
\eea
and obtain Eq.~\eqref{eqn:massshift-fermion-boson} 
for $\Delta m_{A}(L)$.

\subsection{Vertex functions}

\begin{figure}[t]
\centering\includegraphics[width=12cm]{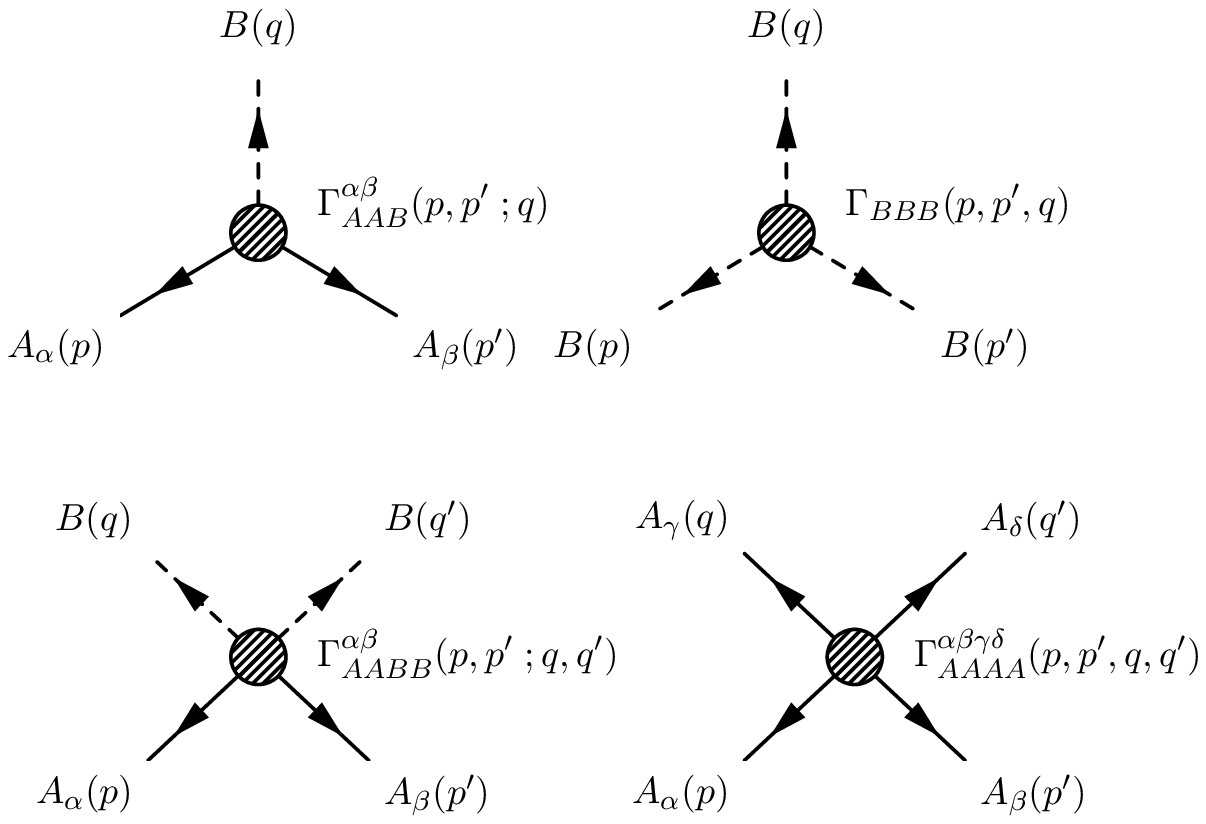}
    \caption{Vertex functions used in
Figs.~\ref{fig:selfenergy}$\sim$\ref{fig:aabar-fermion-scattering}.}
\label{fig:vertex}
\end{figure}

In Fig.~\ref{fig:vertex}, we list all types of 
the vertex functions used in 
Figs.~\ref{fig:selfenergy}$\sim$\ref{fig:aabar-fermion-scattering}.
Arrows represent the direction of momentum flow.
Outgoing momenta from the vertex are parametrized 
to have positive sign.
Greek letters denote spinor indices
for the $\Psi_{A}$-$\phi_{B}$ system.

\section{Ingredients of the formulae}
\label{sec:summary-result}

We list all contributions to $\Sigma_{L}(p)-\Sigma(p)$
in the $\phi_{A}$-$\phi_{B}$ system (Fig.~\ref{fig:selfenergy})
and 
$\bar{u}(p,r)[\Sigma_{L}(p)-\Sigma(p)]u(p,r)$
in the $\Psi_{A}$-$\phi_{B}$ system 
(Fig.~\ref{fig:selfenergy-fermion}).

\subsection{$\phi_{A}$-$\phi_{B}$ system}

\bea
I_{a1} &= &
\frac{3}{4 \pi L}\int_{-\infty}^{\infty} \frac{dq_{0}}{2\pi}
e^{-L\sqrt{q_{0}^{2}+m_{A}^{2}}} \; F_{\mathit{AA}}^{(a1)}(iq_{0})
 +O(e^{-L\bar{m}})
\; ,\\
&&
F_{\mathit{AA}}^{(a1)}(iq_{0})
\equiv 
\Gamma_{\mathit{AAAA}}(-p,p,-q,q) 
|_{p^{2}=q^{2}=-m_{A}^{2}}\; ,
\nonumber\\
&&\nonumber\\
I_{a2} &= &
\frac{3 }{4 \pi L}\int_{-\infty}^{\infty}\frac{dq_{0}}{2\pi}
e^{-L\sqrt{q_{0}^{2}+m_{B}^{2}}}\;  F_{\mathit{AB}}^{(a2)}(iq_{0})
 +O(e^{-L\bar{m}})
\; ,\\
&&
F_{\mathit{AB}}^{(a2)} (iq_{0})\equiv  
\Gamma_{\mathit{AABB}}(-p,p;-q,q)
|_{p^{2}=-m_{A}^{2},\; q^{2}=-m_{B}^{2}}\; ,
\nonumber\\
&&\nonumber\\
I_{b1} &= &
\frac{3 }{4\pi L}
\Biggl [
\frac{\lambda_{\mathit{AAB}}^{2}}{2 \nu_{B}}
e^{-L \sqrt{m_{B}^{2}-\nu_{B}^{2}}}
\nonumber\\*
&&
+ \! \int_{-\infty}^{\infty} \! \frac{dq_{0}}{2\pi} 
e^{-L\sqrt{q_{0}^{2}+m_{A}^{2}}}
 \{ F_{\mathit{AA}}^{(b1-1)}(iq_{0}) 
+F_{\mathit{AA}}^{(b1-2)}(iq_{0}) \}
\nonumber\\*
&&
+\! \int_{-\infty}^{\infty} \! \frac{dq_{0}}{2\pi} 
e^{-L\sqrt{q_{0}^{2}+m_{B}^{2}}}
 \{ F_{\mathit{AB}}^{(b1-1)}(iq_{0}) 
+F_{\mathit{AB}}^{(b1-2)}(iq_{0}) \}
\Biggr ] \! +O(e^{-L\bar{m}}) \; ,\quad \quad\quad
\\
&&
F_{\mathit{AA}}^{(b1-1)} (iq_{0}) 
\equiv 
\Gamma_{\mathit{AAB}}(-p,-q; p+q) G_{B}(p+q)
\nonumber\\*
&&
\quad\quad\quad\quad\quad\quad\quad
\times 
\Gamma_{\mathit{AAB}}(p,q; -p-q)
|_{p^{2}=q^{2}=-m_{A}^{2}} \; ,
\nonumber\\
&&
F_{\mathit{AA}}^{(b1-2)} (iq_{0}) 
\equiv 
\Gamma_{\mathit{AAB}}(-p,q;p-q)G_{B}(p-q)
\nonumber\\*
&&
\quad\quad\quad\quad\quad\quad\quad
\times 
\Gamma_{\mathit{AAB}}(p,-q;-p+q)
|_{p^{2}=q^{2}=-m_{A}^{2}} \; ,
\nonumber\\
&&F_{\mathit{AB}}^{(b1-1)}(iq_{0})
\equiv
 \Gamma_{\mathit{AAB}}(-p,p+q;-q) G_{A}(p+q)
\nonumber\\*
&&
\quad\quad\quad\quad\quad\quad\quad
\times 
\Gamma_{\mathit{AAB}}(p,-p-q;q) 
|_{p^{2}=-m_{A}^{2},\; q^{2}=-m_{B}^{2}} \; ,
\nonumber\\
&&F_{\mathit{AB}}^{(b1-2)}(iq_{0})
\equiv
 \Gamma_{\mathit{AAB}}(-p,p-q;q) G_{A}(p-q)
\nonumber\\*
&&
\quad\quad\quad\quad\quad\quad\quad
\times 
\Gamma_{\mathit{AAB}}(p,-p+q; -q) 
|_{p^{2}=-m_{A}^{2},\; q^{2}=-m_{B}^{2}} \; ,
\nonumber\\
&&\nonumber\\
I_{c1}
&=& 
\frac{3 }{4 \pi L}\int_{-\infty}^{\infty} \frac{dq_{0}}{2\pi}
e^{-L\sqrt{q_{0}^{2}+m_{A}^{2}}} \;
F_{\mathit{AA}}^{(c1)}(iq_{0})   +O(e^{-L\bar{m}}) \; ,\\
&&F_{\mathit{AA}}^{(c1)} (iq_{0}) \equiv  
 \Gamma_{\mathit{AAB}}(-p,p;0)G_{B}(0)
\Gamma_{\mathit{AAB}}(-q,q;0) 
|_{p^{2}=q^{2}=-m_{A}^{2}}\; ,
\nonumber\\
&&\nonumber\\
I_{c2}
&=& 
\frac{3}{4 \pi L}\int_{-\infty}^{\infty} \frac{dq_{0}}{2\pi}
e^{-L\sqrt{q_{0}^{2}+m_{B}^{2}}} \; 
F_{\mathit{AB}}^{(c2)} (iq_{0})  +O(e^{-L\bar{m}})\; ,\\
&&
F_{\mathit{AB}}^{(c2)} (iq_{0}) \equiv 
\Gamma_{\mathit{AAB}}(-p,p;0) G_{B}(0) 
\Gamma_{\mathit{BBB}}(0,-q,q)
|_{p^{2}=-m_{A}^{2},\; q^{2}=-m_{B}^{2}}\; .
 \nonumber
\eea
The forward scattering amplitudes are defined by
\bea
&&
F_{\mathit{AA}}(iq_{0})=
F_{\mathit{AA}}^{(a1)} (iq_{0})+
F_{\mathit{AA}}^{(b1-1)}(iq_{0}) +F_{\mathit{AA}}^{(b1-2)}(iq_{0}) 
+ F_{\mathit{AA}}^{(c1)}(iq_{0})\; ,
\\
&&
F_{\mathit{AB}}(iq_{0})=
F_{\mathit{AB}}^{(a2)} (iq_{0})+
F_{\mathit{AB}}^{(b1-1)}(iq_{0}) +F_{\mathit{AB}}^{(b1-2)}(iq_{0}) 
+F_{\mathit{AB}}^{(c2)}(iq_{0})\; .
\eea

\subsection{$\Psi_{A}$-$\phi_{B}$ system}

\bea
I_{a1} &= &
 - \frac{3}{4 \pi L}\int_{-\infty}^{\infty}\frac{dq_{0}}{2\pi}
e^{-L\sqrt{q_{0}^{2}+m_{A}^{2}}} 
\{ F_{\mathit{AA}}^{(a1)}(iq_{0})
+F_{\mathit{A\bar{A}}}^{(a1)} (iq_{0})\} 
\! +O(e^{-L\bar{m}})
\; ,\\
&&
F_{\mathit{AA}}^{(a1)} (iq_{0})\equiv
-
\sum_{s}^{}
\bar{u}_{\alpha}(p,r)u_{\beta}(p,r) 
\bar{u}_{\gamma}(q,s)u_{\delta}(q,s)
\nonumber\\*
&&
\quad\quad\quad\quad\quad\quad\quad
\times 
 \Gamma_{\mathit{AAAA}}^{\alpha\beta\gamma\delta}
 (-p,p,-q,q) 
|_{p^{2}=q^{2}=-m_{A}^{2}}\; , 
\nonumber\\
&&
F_{\mathit{A\bar{A}}}^{(a1)} (iq_{0})\equiv
\sum_{s}^{}
\bar{u}_{\alpha}(p,r)v_{\beta}(p,r) 
\bar{v}_{\gamma}(q,s)u_{\delta}(q,s)
\nonumber\\*
&&
\quad\quad\quad\quad\quad\quad\quad
\times 
\Gamma_{\mathit{AAAA}}^{\alpha\beta\gamma\delta}(-p,p,q,-q)
|_{p^{2}=q^{2}=-m_{A}^{2}}\; , 
\nonumber\\
&&\nonumber\\
I_{a2} &= &
\frac{3}{4 \pi L}\int_{-\infty}^{\infty}\frac{dq_{0}}{2\pi}
e^{-L\sqrt{q_{0}^{2}+m_{B}^{2}}}\;  F_{\mathit{AB}}^{(a2)} (iq_{0})
 +O(e^{-L\bar{m}}) \; , \\
&&
F_{\mathit{AB}}^{(a2)} \equiv  
\bar{u}_{\alpha}(p,r)
\Gamma_{\mathit{AABB}}^{\alpha\beta}(-p,p;-q,q) 
u_{\beta}(p,r) |_{p^{2}=-m_{A}^{2},\; q^{2}=-m_{B}^{2}}\; ,
\nonumber\\
&&\nonumber\\
I_{b1} &= &
\frac{3 }{4\pi L}
\Biggl [
\frac{\lambda_{\mathit{AAB}}^{2}}{2 \nu_{B}}
e^{-L\sqrt{m_{B}^{2}-\nu_{B}^{2}}}
\nonumber\\*
&&
-  \int_{-\infty}^{\infty}\frac{dq_{0}}{2\pi} 
e^{-L\sqrt{q_{0}^{2}+m_{A}^{2}}}
\; \{ F_{\mathit{AA}}^{(b1)}(iq_{0}) 
+ F_{\mathit{A\bar{A}}}^{(b1)}(iq_{0}) \}
\nonumber\\*
&&
+ \int_{-\infty}^{\infty}\frac{dq_{0}}{2\pi} 
e^{-L\sqrt{q_{0}^{2}+m_{B}^{2}}}
\; \{ F_{\mathit{AB}}^{(b1-1)}(iq_{0}) 
+F_{\mathit{AB}}^{(b1-2)}(iq_{0}) \}
\Biggr ]  +O(e^{-L\bar{m}}) \; , 
\\
&&
F_{\mathit{AA}}^{(b1)} (iq_{0}) 
\equiv 
- 
\sum_{s}^{}
\bar{u}_{\alpha}(p,r)
\Gamma_{\mathit{AAB}}^{\alpha\beta}(-p,-q; p+q)
u_{\beta}(q,s) G_{B}(p+q)
\nonumber\\*
&&
\qquad\qquad\qquad \times 
 \bar{u}_{\gamma}(q,s)
\Gamma_{\mathit{AAB}}^{\gamma\delta}(p,q; -p-q) 
u_{\delta}(p,r) |_{p^{2}=q^{2}=-m_{A}^{2}} \; ,
\nonumber\\
&&
F_{A\bar{A}}^{(b1)} (iq_{0}) 
\equiv 
  \sum_{s}
\bar{u}_{\alpha}(p,r)\Gamma_{\mathit{AAB}}^{\alpha\beta}
(-p,q;p-q) v_{\beta}(q,s) G_{B}(p-q)
\nonumber\\*
&&
\qquad\qquad\qquad \times 
\bar{v}_{\gamma}(q,s) \Gamma_{\mathit{AAB}}^{\gamma\delta}
(p,-q;-p+q) u_{\delta}(p,r) |_{p^{2}=q^{2}=-m_{A}^{2}} \; ,
\nonumber\\
&&
F_{\mathit{AB}}^{(b1-1)}(iq_{0})
\equiv
 \bar{u}_{\alpha}(p,r)\Gamma_{\mathit{AAB}}^{\alpha\beta}
(-p,p+q;-q) S_{A}^{\beta\gamma}(p+q)
\nonumber\\*
&&
\qquad\qquad\qquad \times 
\Gamma_{\mathit{AAB}}^{\gamma\delta}
(p,-p-q;q)u_{\delta}(p,r) 
|_{p^{2}=-m_{A}^{2},\; q^{2}=-m_{B}^{2}} \; ,
\nonumber\\
&&
F_{\mathit{AB}}^{(b1-2)}(iq_{0})
\equiv
\bar{u}_{\alpha}(p,r)\Gamma_{\mathit{AAB}}^{\alpha\beta}
(-p,p-q;q) S_{A}^{\beta\gamma}(p-q) \nonumber\\*
&&
\qquad\qquad\qquad \times 
\Gamma_{\mathit{AAB}}^{\gamma\delta}
(p,-p+q; -q)u_{\delta}(p,r) 
|_{p^{2}=-m_{A}^{2},\; q^{2}=-m_{B}^{2}} \; ,
\nonumber\\
&&\nonumber\\
I_{c1}
&=& 
-  \frac{3}{4 \pi L}\int_{-\infty}^{\infty}\frac{dq_{0}}{2\pi}
e^{-L\sqrt{q_{0}^{2}+m_{A}^{2}}} \;
\{ F_{\mathit{AA}}^{(c1)}(iq_{0}) 
+  F_{\mathit{A\bar{A}}}^{(c1)}(iq_{0}) \} +O(e^{-L\bar{m}})\; ,
\quad \quad\quad
\\
&&
F_{\mathit{AA}}^{(c1)} (iq_{0}) \equiv  
- 
\sum_{s}
\bar{u}_{\alpha}(p,r)\Gamma_{\mathit{AAB}}^{\alpha\beta}
(-p,p;0) u_{\beta}(q,s) 
\nonumber\\*
&&
\qquad\qquad\qquad \times 
G_{B}(0)
\bar{u}_{\gamma}(q,s)\Gamma_{\mathit{AAB}}^{\gamma\delta}
(-q,q;0) u_{\delta}(p,r) 
|_{p^{2}=q^{2}=-m_{A}^{2}}\; ,
\nonumber\\
&&
F_{\mathit{A\bar{A}}}^{(c1)} (iq_{0}) \equiv  
 \bar{u}_{\alpha}(p,r) \Gamma_{\mathit{AAB}}^{\alpha\beta}
(-p,p;0) v_{\beta}(q,s)\nonumber\\*
&&
\qquad\qquad\qquad \times 
G_{B}(0)
\bar{v}_{\gamma}(q,s) \Gamma_{\mathit{AAB}}^{\gamma\delta}
(-q,q;0) u_{\delta}(p,r) 
|_{p^{2}=q^{2}=-m_{A}^{2}}\; ,
\nonumber\\
&&\nonumber\\
I_{c2}
&=& 
\frac{3}{4 \pi L}\int_{-\infty}^{\infty}\frac{dq_{0}}{2\pi}
e^{-L\sqrt{q_{0}^{2}+m_{B}^{2}}} \; 
F_{\mathit{AB}}^{(c2)} (iq_{0})  +O(e^{-L\bar{m}}) \; ,
\\
&&
F_{\mathit{AB}}^{(c2)} (iq_{0}) \equiv 
\bar{u}_{\alpha}(p,r)\Gamma_{\mathit{AAB}}^{\alpha\beta}
(-p,p;0) u_{\beta}(p,r) G_{B}(0) 
\nonumber\\*
&&
\quad\quad\quad\quad\quad\quad
\times 
\Gamma_{\mathit{BBB}}(0,-q,q)
|_{p^{2}=-m_{A}^{2},\; q^{2}=-m_{B}^{2}}\; .
\nonumber
\eea
The forward scattering amplitudes are defined by
\bea
&&
F_{\mathit{AA}}(iq_{0})=
F_{\mathit{AA}}^{(a1)} (iq_{0})+
F_{\mathit{AA}}^{(b1)}(iq_{0}) 
+ F_{\mathit{AA}}^{(c1)}(iq_{0})\; ,
\\
&&
F_{\mathit{A\bar{A}}}(iq_{0})=
F_{\mathit{A\bar{A}}}^{(a1)}(iq_{0}) +
F_{\mathit{A\bar{A}}}^{(b1)}(iq_{0})
+ F_{\mathit{A\bar{A}}}^{(c1)}(iq_{0})\; ,
\\
&&
F_{\mathit{AB}}(iq_{0})=
F_{\mathit{AB}}^{(a2)}(iq_{0}) +
F_{\mathit{AB}}^{(b1-1)}(iq_{0}) +F_{\mathit{AB}}^{(b1-2)}(iq_{0}) 
+F_{\mathit{AB}}^{(c2)}(iq_{0})\; .
\eea

\end{document}